\documentclass{nature}

\usepackage{graphicx}
\usepackage{epsfig}
\usepackage[font={small}]{caption}
\usepackage{tablefootnote}
\usepackage{hyperref}
\usepackage[dvipsnames]{xcolor}
\usepackage{comment}

\def\gtorder{\mathrel{\raise.3ex\hbox{$>$}\mkern-14mu
             \lower0.6ex\hbox{$\sim$}}}
\def\ltorder{\mathrel{\raise.3ex\hbox{$<$}\mkern-14mu
             \lower0.6ex\hbox{$\sim$}}}

\usepackage{amssymb}
\usepackage{amsmath}
\usepackage{tablefootnote}

\newcommand{\aap}{Astron. Astrophys.}
\newcommand{\araa}{Annu. Rev. Astron. Astrophys.} 
\newcommand{\apj}{Astrophys. J.}
\newcommand{\aj}{Astron. J.}
\newcommand{\apjl}{Astrophys. J. Lett.}
\newcommand{\apjs}{Astrophys. J. Suppl. Ser.}
\newcommand{\nat}{Nature}
\newcommand{\pasp}{Publ. Astron. Soc. Pac.}

\newcommand{\mnras}{Mon. Not. R. Astron. Soc.}

\newcommand{\ssr}{Space Sci. Rev.}  
\newcommand{\nastro}{Nat. Astron.}

\title{An asymmetric electron-scattering photosphere around optical tidal disruption events}

\author{Giorgos Leloudas$^{1,*}$, Mattia Bulla$^{2}$, Aleksandar Cikota$^{3}$, Lixin Dai$^{4,*}$, Lars L. Thomsen$^{4}$, Justyn R. Maund$^{5}$, Panos Charalampopoulos$^{1}$,  Nathaniel Roth$^{6}$, Iair Arcavi$^{7,8}$, Katie Auchettl$^{9,10,11}$,
Daniele B. Malesani$^{1,12}$,
Matt Nicholl$^{13}$,
Enrico Ramirez-Ruiz$^{11}$}
\begin{document}

\maketitle

\begin{affiliations}
   \item DTU Space, National Space Institute, Technical University of Denmark, Elektrovej 327, 2800 Kgs. Lyngby, Denmark
   \item The Oskar Klein Centre, Department of Astronomy, Stockholm University, AlbaNova, SE-106 91, Stockholm, Sweden
   \item European Organisation for Astronomical Research in the Southern Hemisphere (ESO), Alonso de Cordova 3107, Vitacura, Casilla 19001, Santiago de Chile, Chile
   \item Department of Physics, The University of Hong Kong, Pokfulam Road, Hong Kong
   \item Department of Physics and Astronomy, The University of Sheffield, Hicks Building, Hounsfield Road, Sheffield, S3 7RH, UK
    \item Lawrence Livermore National Laboratory, Livermore, CA 94550, USA    
    \item The School of Physics and Astronomy, Tel Aviv University, Tel Aviv 69978, Israel
    \item CIFAR Azrieli Global Scholars program, CIFAR, Toronto, Canada
    \item{School of Physics, The University of Melbourne, VIC 3010, Australia}
    \item{ARC Centre of Excellence for All Sky Astrophysics in 3 Dimensions (ASTRO 3D)}
    \item{Department of Astronomy and Astrophysics, University of California, Santa Cruz, CA 95064, USA}
    \item Department of Astrophysics/IMAPP, Radboud University, PO Box 9010,
    6500 GL, The Netherlands
    \item Birmingham Institute for Gravitational Wave Astronomy and School of Physics and Astronomy, University of Birmingham, Birmingham B15 2TT, UK  
    
$^*$ Corresponding authors
\end{affiliations}


\clearpage

\begin{abstract}
A star crossing the tidal radius of a supermassive black hole  will be spectacularly ripped apart with an accompanying burst of radiation. A few tens of such tidal disruption events (TDEs) have now been identified in optical wavelengths, but the exact origin of the strong optical emission remains inconclusive. Here we report polarimetric observations of three TDEs. The continuum polarization appears independent of wavelength, while emission lines are partially depolarized. These signatures are consistent with photons being scattered and polarized in an envelope of free electrons. An almost axisymmetric photosphere viewed from different angles is in broad agreement with the data, but there is also evidence for deviations from axial symmetry before the peak of the flare and significant time evolution at early times, compatible with the rapid formation of an accretion disk. By combining a super-Eddington accretion model with a radiative transfer code we simulate the polarization degree as a function of disk mass and viewing angle, and we show that the predicted levels are compatible with the observations, for extended reprocessing envelopes  of $\sim$1000 gravitational radii. Spectropolarimetry therefore constitutes a new observational test for TDE models, and opens an important new line of exploration in the study of TDEs. 
\end{abstract}


The polarization of light can provide a direct probe of the photosphere geometry and shed light on the physical mechanisms powering transient phenomena\cite{WangWheeler,BullaKN,Maund2020,Wiersema2012} and active galactic nuclei (AGN)\cite{AntonucciMiller1985,Smith2004}. 
Spectropolarimetry, however, is a photon-hungry technique and requires bright targets. 
Our sample therefore comprises some of the brightest and most nearby TDEs observed after 2018. 
AT~2018dyb (a.k.a. ASASSN-18pg) is a well-observed, nearby ($z = 0.018$) TDE that showed Bowen fluorescence lines in its spectrum\cite{Leloudas2019,Holoien2020}.
We observed AT~2018dyb with the the Very Large Telescope (VLT) on four epochs, between $-17$ and $+180$ days (all phases are quoted in the rest frame and with respect to the peak\cite{Leloudas2019} of the light curve), including two epochs of spectropolarimetry. 
Our second target was AT~2019azh, a TDE dominated by broad Balmer lines at $z = 0.022$ (a.k.a. ASASSN-19dj). 
Our observations were conducted at $+22$ days when the TDE showed weak X-ray emission ($L_X \sim 10^{41}$ erg s$^{-1}$). Remarkably, the X-rays brightened by one order of magnitude 200 days later, followed by a flare in the radio\cite{Hinkle2020,Wevers2020,vanvelzen2021,sfaradi2019azh}. 
Finally, we obtained data for AT~2019dsg, a TDE possibly associated with a high-energy neutrino\cite{Stein2020}, and showing narrow emission lines in its spectrum\cite{Cannizzaro2021} at $z = 0.052$. 
X-rays at a level of $L_X \sim 2.5 \times 10^{43}$~erg~s$^{-1}$ were detected from AT~2019dsg at early phases but faded rapidly, while the source also showed persistent radio emission.
Our VLT spectropolarimetry at $+33$ days was complemented by broad-band polarimetry from the Nordic Optical Telescope (NOT), including both published\cite{Lee2020} and our own unpublished data.
A detailed observing log summarising all data is given in Table~\ref{tab:log} and the polarization results are presented in Table~\ref{tab:quP}. 

\noindent  \textbf{Insights from observations}
 
\begin{figure}
\centering
\includegraphics[width=\textwidth]{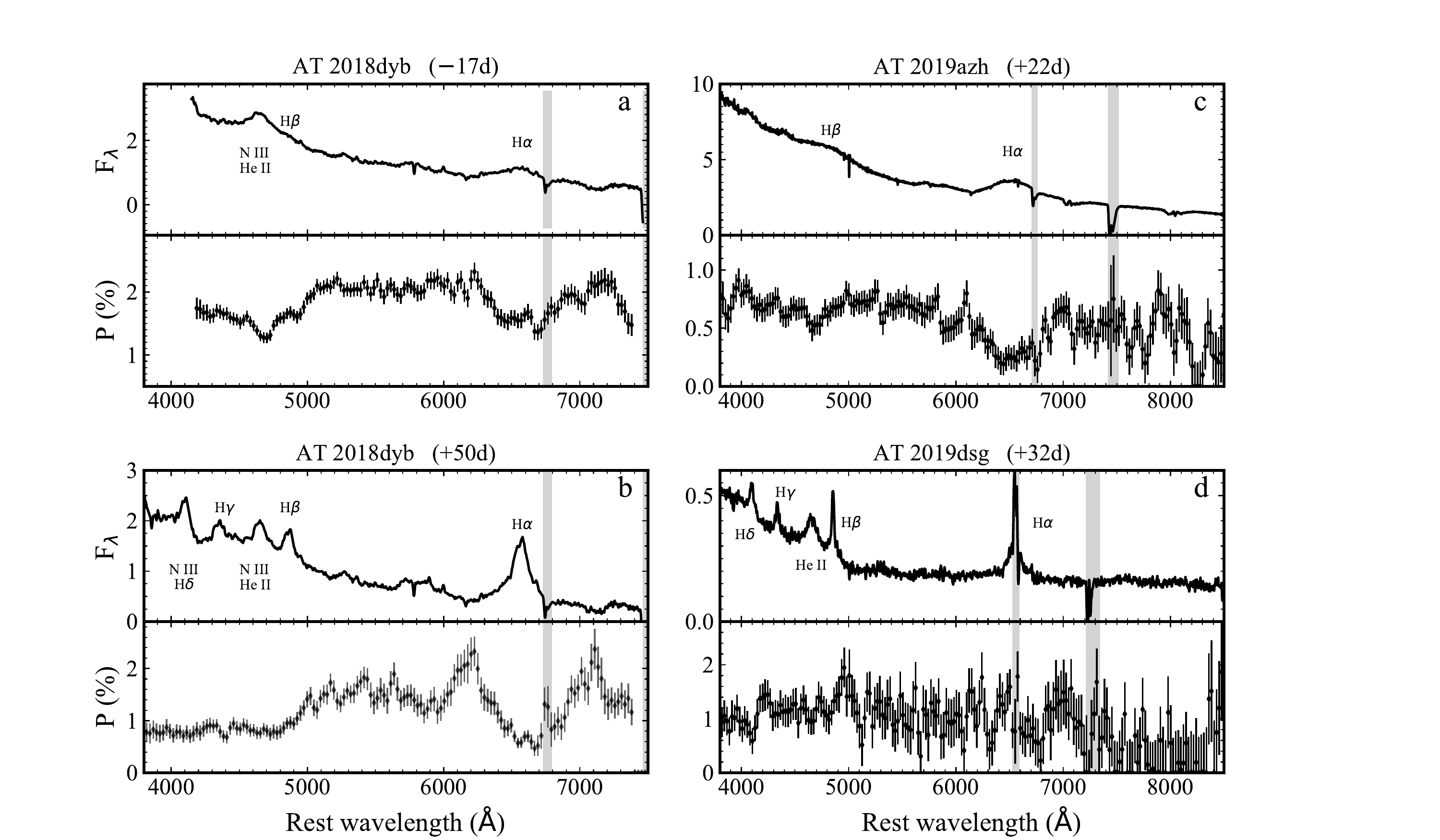}
\caption{\boldmath$\vert$\unboldmath \hspace{1em} \textbf{Spectral polarimetry of optical TDEs.} 
The flux spectra  are shown in the top panels, while polarization degree is shown in the bottom panels as a function of wavelength. All error bars are 1$\sigma$ uncertainties and F$_\lambda$ is in units of erg s$^{-1}$ cm$^{-2}$ $\mathrm{\AA}^{-1}$.
The TDE names and observation phases are indicated above the panels. Prominent emission lines are labelled on the spectra and regions of telluric absorption are indicated with gray bands.
The continuum polarization appears constant with wavelength in line-free regions (e.g. between 5000 -- 6000 \AA). Depolarization occurs at the location of broad emission lines. 
All data is shown here after correcting for the ISP and for the host dilution.
\label{fig:flux_pola_spec}}
\end{figure}

The flux and polarization spectra of the three TDEs (two epochs for AT~2018dyb) are shown in Figure~\ref{fig:flux_pola_spec}, and the Stokes parameters and polarization angle are additionally shown in Extended Data Figure~\ref{fig:resultsall_cor}. 
During the analysis, these spectra have been subject to two important corrections for 
the effect of insterstellar polarization (ISP), i.e. polarization induced by dust grains along the line of sight, and for the effect of dilution by unpolarized light from the TDE host galaxy.
TDEs are buried in the nucleus of their hosts, and to study their intrinsic polarization the light contribution from the host needs to be removed. This can be complicated, as the host contribution evolves with time and has a strong wavelength dependence typically increasing towards longer wavelengths (Extended Data Figure~\ref{fig:HostContrib}), because TDEs are typically blue and their host galaxies red. 
Compared with AGNs, however, where this correction has also posed significant challenges\cite{MillerAntonucci1983,Marin2018}, TDEs eventually fade away allowing for a more accurate removal of the host. 
We have carefully applied these non-trivial but critical corrections following the procedures in the Methods section but we also present the fully reduced data before corrections in Extended Data Figures~\ref{fig:flux_pola_spec_orig} and~\ref{fig:resultsall_orig}.
Figure~\ref{fig:flux_pola_spec} shows that TDEs have a polarization level that is overall constant in regions that are continuum-dominated and free of strong emission lines (e.g. between 5000 -- 6000 \AA).
The measured continuum polarization levels are moderate, ranging from 0.7\% in AT~2019azh to 1.5 -- 2.1\% in AT~2018dyb.

\begin{figure}
\centering
\includegraphics[width=16cm]{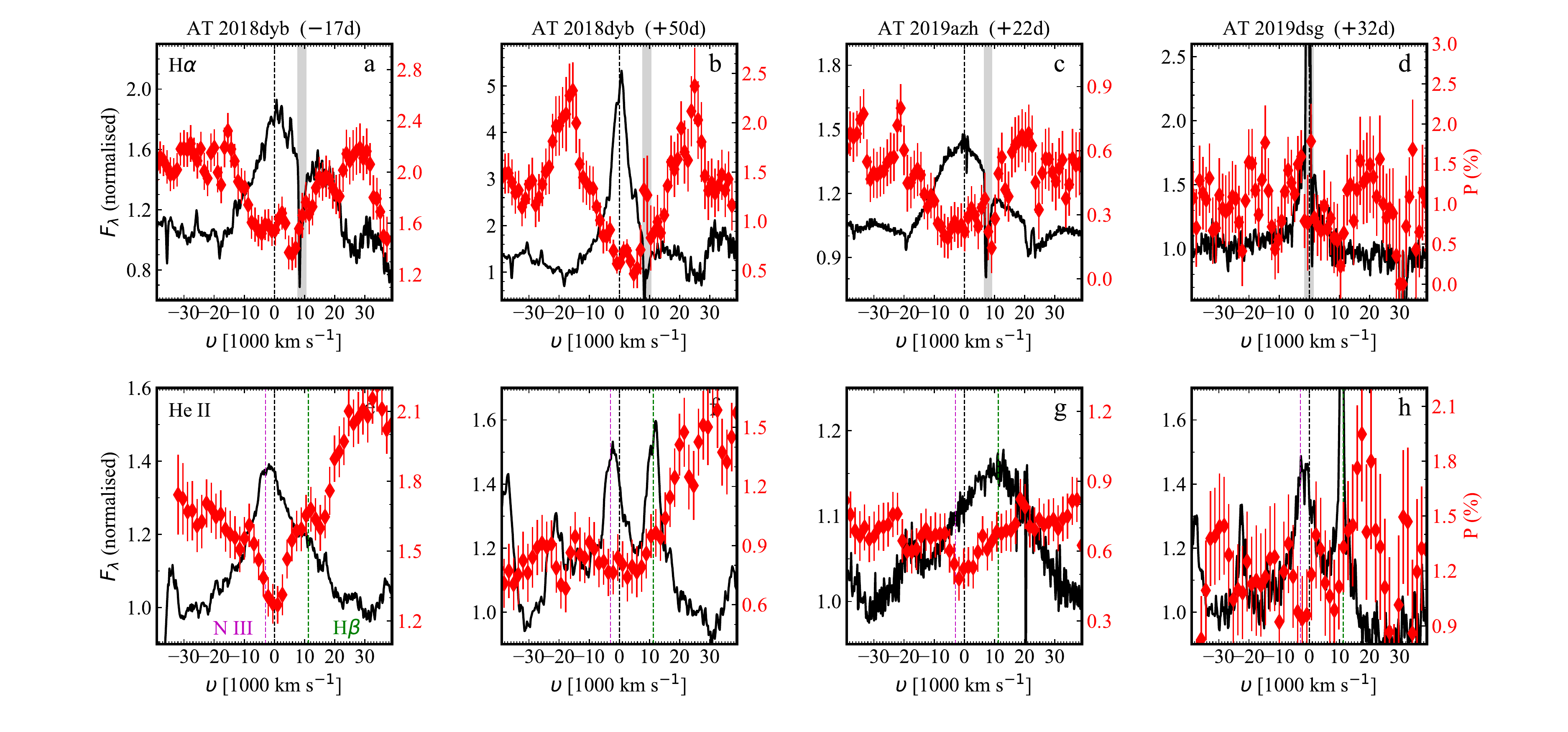}
\caption{\boldmath$\vert$\unboldmath \hspace{1em} \textbf{Emission line and polarization profiles.} The first row shows the polarization (red, right y-axis) across the H$\alpha$ emission line  (black, left y-axis), normalised to the continuum and shown in velocity space. 
The TDE names and observation phases are specified above the panels. Regions of telluric absorption are indicated with gray bands.
The second row includes similar graphs but centered at the He II line. Magenta and green dashed lines mark the position of N III and H$\beta$. 
All error bars are 1$\sigma$ uncertainties.
All data is shown here after correcting for the ISP and for the host dilution.
\label{fig:Ha_He_corrected}
}
\end{figure}

At the location of strong emission lines, including H$\alpha$ and the plethora of lines between 4000--5000 \AA, the spectrum is depolarized.
The line profiles and corresponding polarization spectra are plotted in more detail in Figure~\ref{fig:Ha_He_corrected} in velocity space: panels (a) -- (d) show the H$\alpha$ line and panels (e) -- (h) the N III, He II, H$\beta$ complex (and Extended Data Figure~\ref{fig:Ha_He_orig} shows the same data before the ISP and host corrections). 
Focusing on H$\alpha$, we observe
that the core of the emission line is indeed partially depolarized, but the wings of the lines (velocities $\sim$ 20,000 km s$^{-1}$) show polarization peaks, especially prominent for AT~2018dyb and AT~2019azh. 
The profile of AT~2019dsg is more complex, but this event is also different spectroscopically, showing both a broad and a narrow component that does not evolve with time\cite{Cannizzaro2021}.
It is possible that the host of AT~2019dsg contains an AGN, which can contribute in the polarization by a hidden broadlined region, similar to what has been observed in Seyfert 2 galaxies\cite{AntonucciMiller1985}.
Partial depolarization also occurs in the blue part of the TDE spectra and it is stronger in the core of He~II than of H$\beta$.
Interestingly, this is even the case for AT~2019azh although strong He~II does not appear in the flux spectrum at this phase, indicating that the line might be present but subtle radiative transfer effects could make it weak\cite{Roth16}.

\begin{figure}
\centering
\includegraphics[width=16cm]{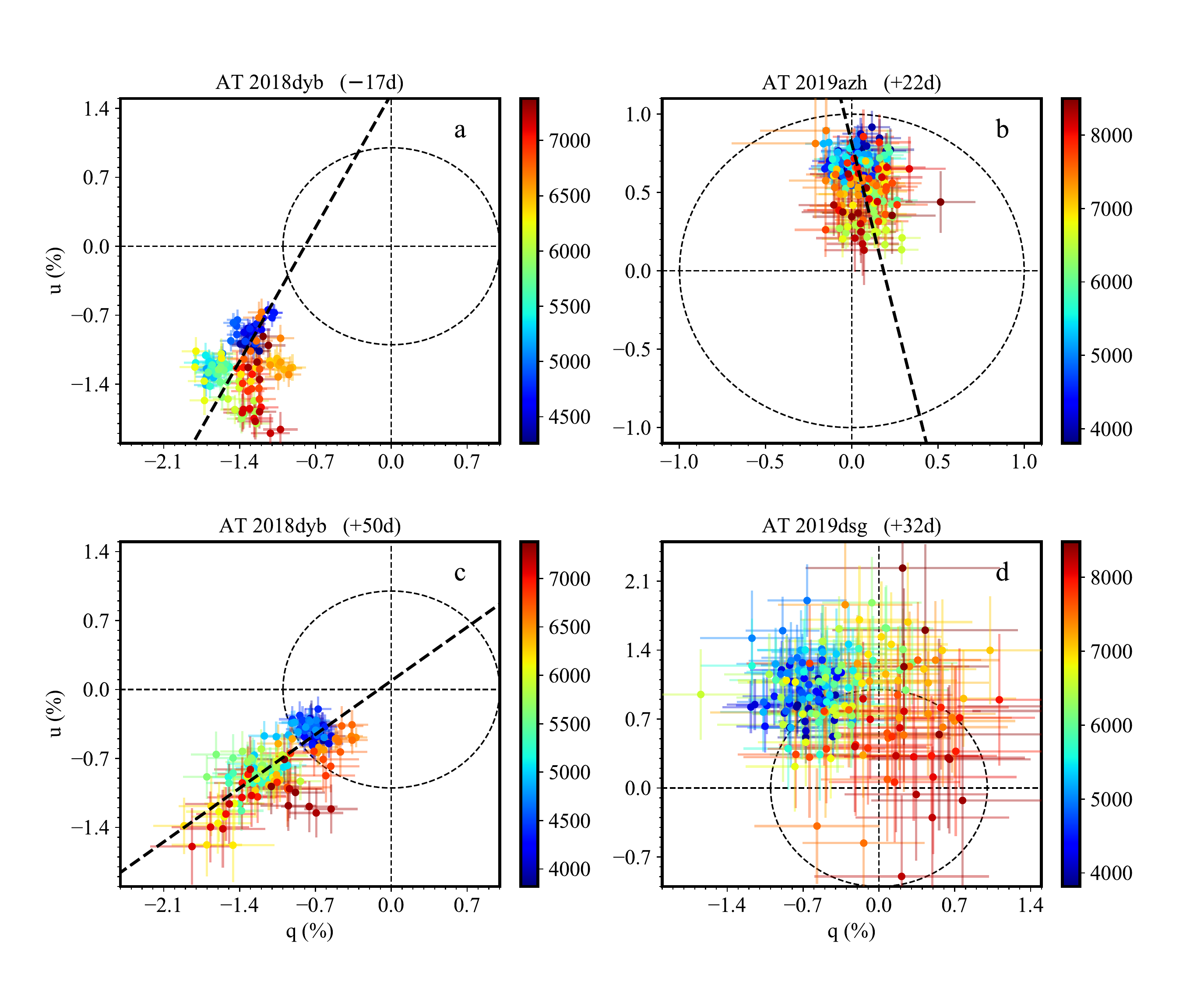}
\caption{\boldmath$\vert$\unboldmath \hspace{1em} \textbf{The Stokes plane.} 
The TDE spectropolarimetric data on the Stokes $q$ -- $u$ plane. Points are colored according to their wavelength, as indicated in the colorbars. A linear fit to the data (dominant axis) is shown as a thick dashed line. 
No reliable fit is possible for AT~2019dsg. 
Thin dashed lines mark $q=0$, $u=0$ and $P=1$\%. 
All error bars are 1$\sigma$ uncertainties.
All data is shown here after correcting for the ISP and for the host dilution.
\label{fig:Stokes_norot_corrected}
}
\end{figure}

Figure~\ref{fig:Stokes_norot_corrected} shows the spectropolarimetric data of the three TDEs on the Stokes $q$ -- $u$ plane, after correcting for the ISP and the host dilution (and Extended Data Figure ~\ref{fig:Stokes_norot_original} shows the data without these corrections). 
For AT~2018dyb, evolution occurs between $-$17 and $+$50 days:  
in the second epoch, the data cloud  appears more structured and a linear fit to the data passes close to the origin. 
The same can be stated for AT~2019azh at $+$22 days.
The data of AT~2019dsg is of lower S/N and the host correction adds additional noise to this dataset. Extended Data Figure~\ref{fig:Stokes_norot_original}, however, shows that AT~2019dsg at $+32$ days can also be fit with an axis that passes close to the origin, before the host correction.
It is possible to rotate the original coordinate system so that the new $q_\mathrm{rot}$ axis becomes parallel to the best-fit axis. 
Similar transformations in the $q$ -- $u$ plane have been widely used in the past, including for supernova explosions\cite{WangWheeler}, where the best-fit axis is sometimes referred to as the dominant axis.
This transformation is equivalent to decomposing the polarimetry in two orthogonal components, one parallel and one orthogonal to the dominant axis, and it has been argued that 
the existence of a dominant axis is an indication of axial symmetry, while scatter and deviations around this axis indicate departures from an axisymmetric geometry\cite{WangWheeler}.
The rotated Stokes planes are shown in Extended Data Figure~\ref{fig:Stokes_rot_corrected}
(no reliable fit and rotation is possible for AT~2019dsg) and the rotated Stokes parameters are plotted against wavelength in Extended Data Figure~\ref{fig:resultsall_cor}.
While post-maximum data are generally close to $u_\mathrm{rot} = 0$, albeit with some dispersion and moderate offsets, this is certainly not the case for the pre-maximum data of AT~2018dyb, which shows a large systematic offset from $u_\mathrm{rot}$. 
We therefore deduce that AT~2018dyb was far from axial symmetry at $-17$ days, but settled to an almost axially symmetric geometry a few weeks later. 

\begin{figure}
\centering
\includegraphics[width=16cm]{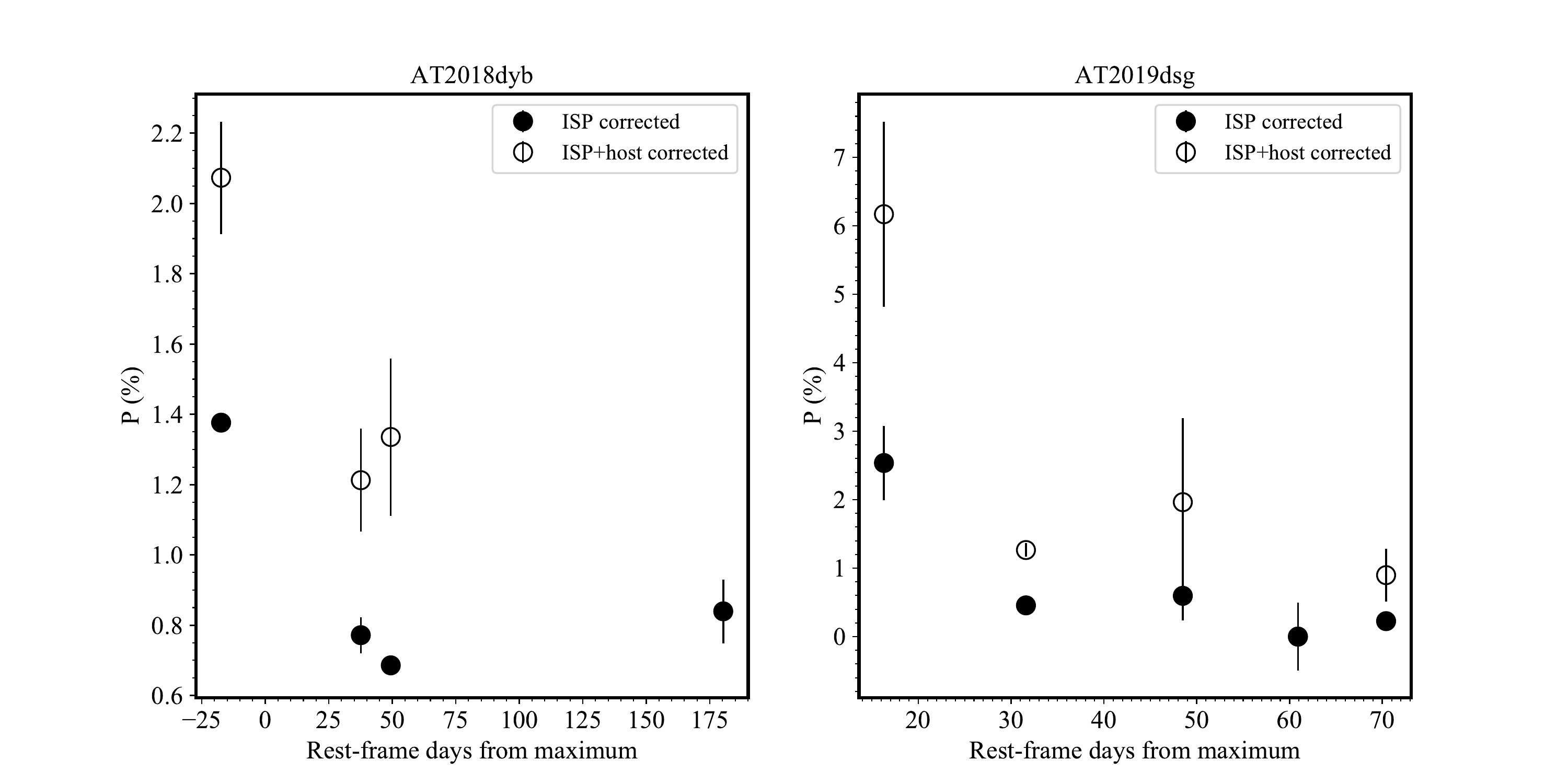}
\caption{\boldmath$\vert$\unboldmath \hspace{1em}  \textbf{
The evolution of polarization with time for two TDEs.}
Measurements are in the $V$ band, which corresponds to a relatively line-free wavelength region, and is thus representative of the continuum.
We plot data corrected for the ISP only (filled circles) and for the ISP+host dilution (empty circles).
All error bars are 1$\sigma$ uncertainties.
The host correction induces additional uncertainties (see Table~\ref{tab:quP}). This makes the last fully-corrected data point  of AT~2018dyb non constraining  and it has therefore not been plotted. For both TDEs, polarization decreases with time. 
\label{fig:timevol}
}
\end{figure}

More information on the time evolution can be obtained by including broad-band polarimetry data, which is available for additional epochs (Table~\ref{tab:quP}). 
This is done in  Figure~\ref{fig:timevol}
for AT~2018dyb and AT~2019dsg. 
We show the evolution in the $V$ band, which covers a relatively line-free region and is thus a better probe for the continuum polarization.
For both TDEs the degree of polarization decreases with time and possibly stabilises a couple of weeks after maximum light. Especially for AT~2019dsg, the time evolution is rather rapid, although this conclusion is mostly based on a single point with large uncertainties\cite{Lee2020} (see Methods for a detailed re-analysis of this data point). 
We hypothesize that the time evolution and convergence to an axisymmetric configuration is related to the formation and circularisation of the accretion disk\cite{Bonnerot21}. 
Independent evidence\cite{Wevers2019} also argues for rapid disk formation in some TDEs. 

\noindent  \textbf{The origin of polarization in optical TDEs}

Our observations suggest that electron scattering is the dominant source of polarization in TDEs.
The electron scattering opacity is wavelength independent and can naturally explain a continuum polarization that is constant with wavelength, as opposed to free-free and bound-free opacities, which are wavelength dependent\cite{Roth16,Inserra2016} (Extended Data Figure~\ref{fig:wavedep_a}).
Synchrotron radiation has been proposed to be responsible for the polarization in relativistic TDEs\cite{Wiersema2012,Wiersema2020}, which are likely viewed through a polar jet, but its contribution to the optical TDEs we study here is shown to be negligible (Methods). Likewise, we demonstrate in Methods that polarization by dust scattering is also subdominant due to the low dust covering factor and the polarization time evolution.
Furthermore, polarization peaks at the wings of emission lines are expected if line broadening in TDEs is caused by electron scattering rather than kinematics\cite{Roth18} and similar signatures have been attributed in this way in supernova explosions\cite{Wang2004,Patat2011}.
Our data therefore directly demonstrates that 
optical radiation is scattered and polarized by free electrons in an outer reprocessing envelope.
The existence of this extended envelope of free electrons is anticipated in models associated with super-Eddington debris fallback and large accretion luminosity\cite{LodatoRossi,Guillochon2014}, while it is unclear whether such high level of ionization would be produced when only stream self-crossings power the  emissions\cite{Piran2015}. The depolarization of the cores of emission lines is a natural prediction of the emission models in which X-rays produced from the inner disk are reprocessed by a large envelope or optically thick wind\cite{Roth16,Roth18,Dai18}, where the emission line photospheres lie outside the optical continuum photosphere.

\noindent \textbf{Models}

\begin{figure}
\centering
\includegraphics[width=0.29\textwidth]{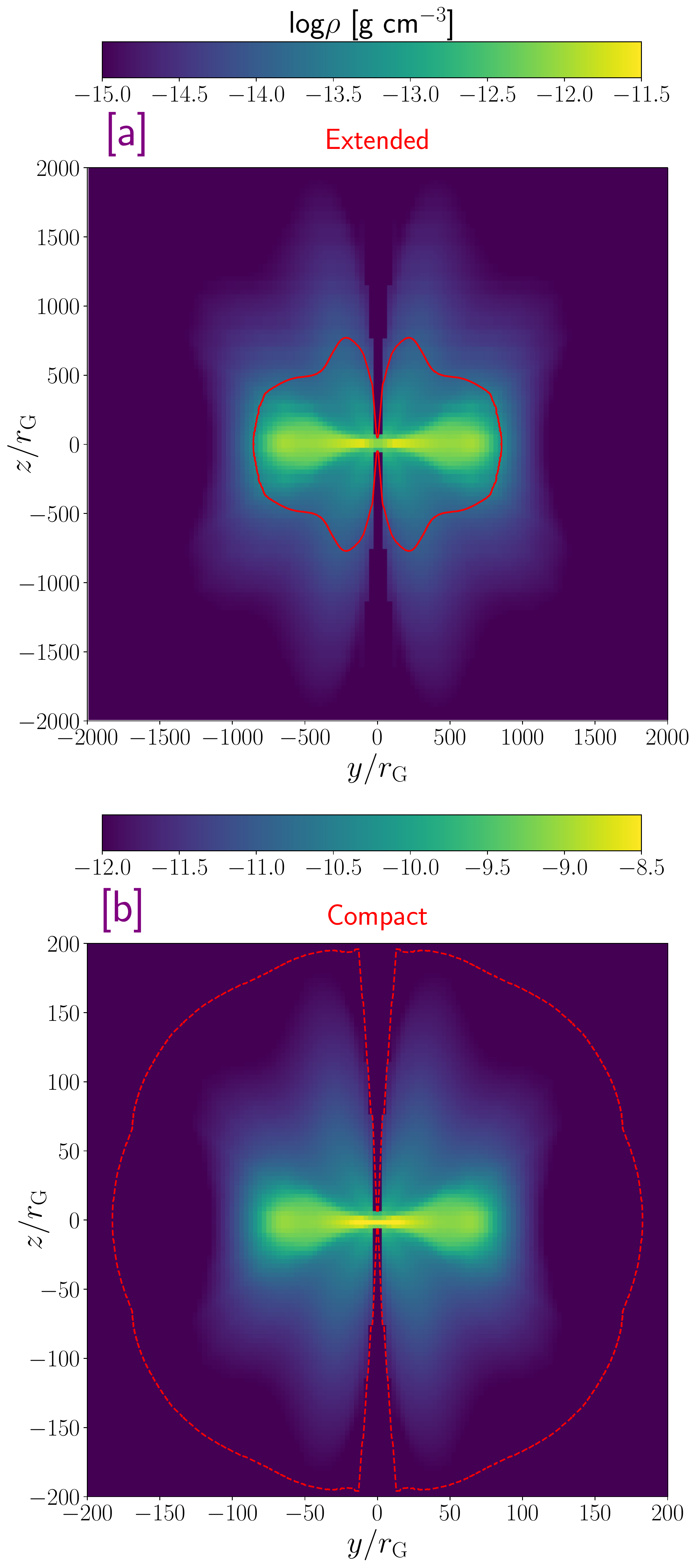}
\includegraphics[width=0.703\textwidth]{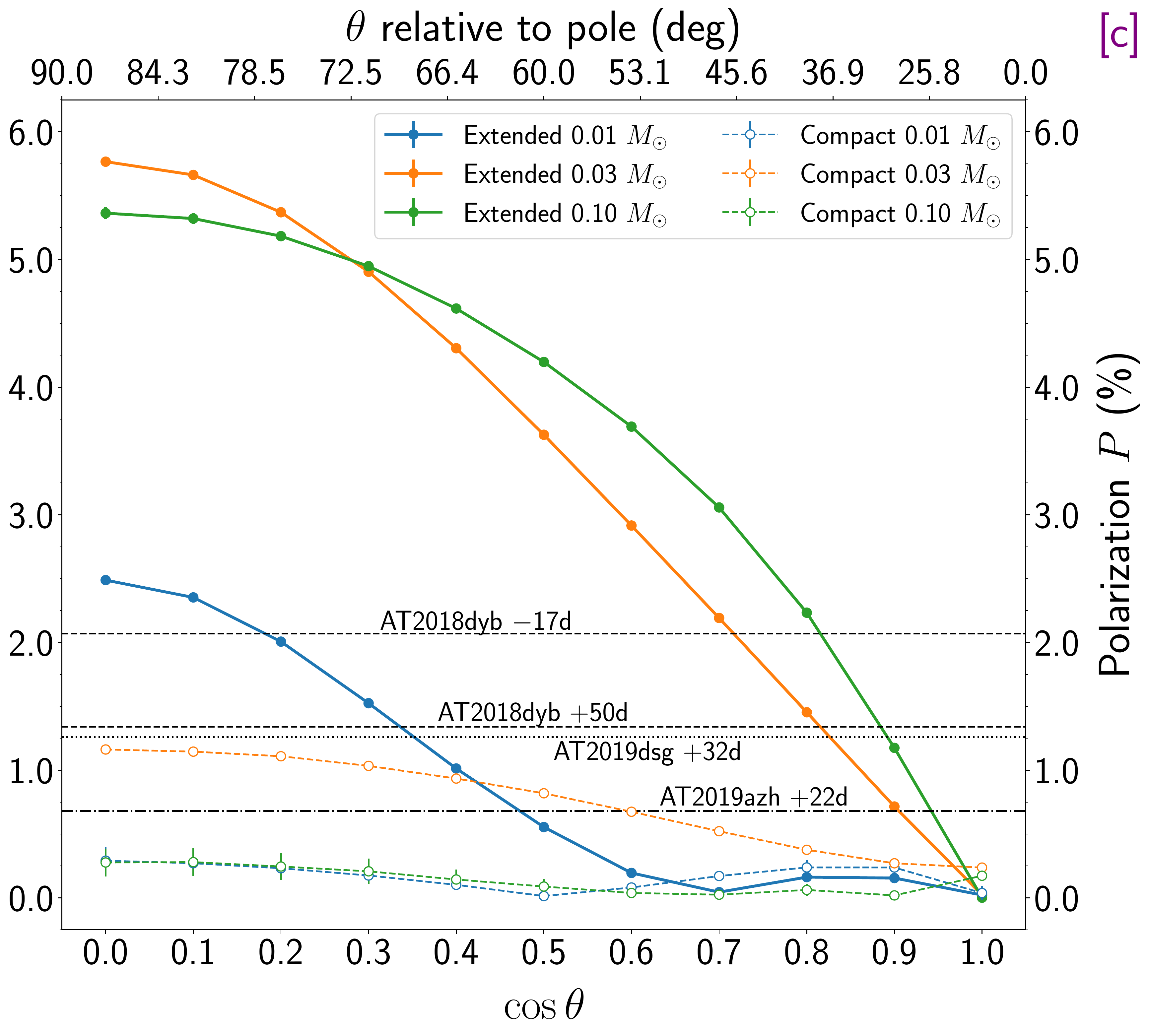}
\caption{\boldmath$\vert$\unboldmath \hspace{1em}
\textbf{Polarization modelling for a TDE super-Eddington disk.} {\it Left panels}: Density distribution of the original disk model\cite{Dai18} with an extended envelope (panel a) and the  compact disk model with the same structure and disk mass as the extended model but with size reduced by a factor of 10 (panel b). Both maps are shown for $m_\mathrm{disk}=0.1\,M_\odot$ and  $M_\mathrm{BH}=5\times10^6\,M_\odot$. The spatial coordinates are given in units of the BH gravitational radius $r_{\rm G}$. The electron scattering photospheres are also overplotted as the red contours. {\it Right panel}: Polarization as a function of viewing angle for the simulated TDE disk assuming three different disk masses $m_\mathrm{disk}$: 0.01 (cyan), 0.03 (orange) and 0.1 (green) $M_\odot$. Filled circles and solid lines refer to the extended models (panel a), while open circles and dashed lines to the compact models (panel b). The observed continuum polarization levels for the three TDEs at the epochs of VLT spectropolarimetry are shown with horizontal lines. 
\label{fig:models_geometry}
}
\end{figure}

The observed polarization properties allow us to put constraints on the geometry of the photosphere, given a specific model. 
We adopted the structure of a simulated TDE super-Eddington disk\cite{Dai18,thomsen2022}
which has been proposed as a unifying model for TDEs (Figure~\ref{fig:models_geometry}(a)), and calculated the predicted polarization level seen from this disk using the Monte Carlo radiative transfer code \textsc{possis}\cite{BullaPOSSIS}.
We carried out a full parametric study (see Methods), and find that the observed polarization level primarily depends on i) the total mass included in the disk $m_\mathrm{disk}$, ii) the compactness of the disk, and iii) the viewing angle.
As the disk mass varies in the range of $0.01-0.1\,M_\odot$, the disk can produce a polarization level of $\sim 1-6\%$ for most viewing angles (Figure~\ref{fig:models_geometry}(c)), which is compatible with the observed values. If we  shrink the disk size by ten times while keeping the same disk mass as shown in Figure~\ref{fig:models_geometry}(b), the resulting compact disks are found to produce polarization mostly below 1\%, which is too low to explain the observed values from AT2018dyb and AT2019dsg. This depolarization effect from
compact disks is due to their higher densities and larger scattering optical depths compared to the
extended disks. 
As the electron scattering optical depth scales as $\tau\sim \rho r \sim m_{\rm disk}/r^2$, it would be possible to obtain the same polarization as the extended model by additionally reducing the disk mass of the compact model by a factor of 100 (i.e. $m_{\rm disk}=0.001\,M_\odot$). 
It is, however, unlikely that such a low-mass disk would be able to efficiently reprocess X-rays, and produce an optical/UV TDE.
Therefore, the observed high level of polarization indicates that the TDE gas flow
should have an extended configuration around the supermassive black hole. For the disk structure
we adopted, the size of the scattering photosphere needs to be around 
1000 black hole gravitational radii $r_G \equiv G M_{\rm BH}/c^2$ (where $G$, $M_{\rm BH}$ and $c$ are the gravitational constant, the black hole mass and the speed of light, respectively), which corresponds to $10^{14}-10^{15}$ cm for black holes of 1-10 million solar masses. This size is consistent with the optical photospheres of  observed TDEs\cite{Gezari2021,vanvelzen2020}. 
In addition, this model can naturally reproduce the declining trend of polarization with time: the disk mass $m_\mathrm{disk}$ is expected to decrease when the debris fallback and accretion rates drop at later times\cite{thomsen2022}, and this leads to a decrease in the polarization for a given viewing angle (Extended Data Figure~\ref{fig:Qmodels}).

\noindent  \textbf{Future outlook}

The polarization properties of TDEs can constrain the geometry and distribution of debris gas around supermassive black holes. Our analysis of the polarization signals from three TDEs show that the 
optical emitting region is aspherical but it becomes consistent
with an almost axisymmetric configuration soon after  the flare peak.
Excitingly, the predicted polarization signal for a super-Eddington accretion disk model shows a clear viewing angle dependence and could potentially be used to constrain the inclination angle for TDEs, especially when combined with additional observables\cite{charalampopoulos2022}.
At present, the degeneracy between viewing angle and disk mass  (Figure~\ref{fig:models_geometry}) 
prevents us from placing firm constraints on the viewing angle and 
independent constraints on $m_{\rm disk}$ will be required. 
Radiative transfer modeling, extended to include a detailed modeling of line polarization and applied  
to alternative TDE models\cite{LodatoRossi, Piran2015,LuBonnerot2020}, combined with future multi-epoch polarimetric observations, have the potential to greatly enhance our understanding of TDEs and map their diversity.


\clearpage

\begin{methods}

This section describes the data, methods  and theoretical calculations used in the main paper.


\subsection{Observations and data reduction}

The data have been acquired using the FOcal Reducer/low dispersion Spectrograph 2 (FORS2)\cite{1998Msngr..94....1A} mounted at the ESO's Very Large Telescope UT1 located on Cerro Paranal in Chile, and the Alhambra Faint Object Spectrograph and Camera (ALFOSC) mounted at the 2.56-m Nordic Optical Telescope (NOT) located at La Palma, Spain.
FORS2 is a dual-beam polarimeter with a Wollaston prism, which splits the incoming beam into two beams with orthogonal directions of polarization (ordinary and extraordinary beam), displaced by $\sim$22$''$. To avoid overlapping of the two beams, a strip mask is inserted in the focal area of FORS2.
ALFOSC uses a half-wave retarder plate (HWP) in the FAPOL unit and a calcite plate mounted in the aperture wheel. 
No mask is used and the ordinary and extraordinary components appear on the same frame, separated by 15$''$.

With FORS2 we obtained observations in both imaging-polarimetry (IPOL) and spectropolarimetry (PMOS) modes, while with ALFOSC we obtained imaging polarimetry only.
All observations were taken with the HWP positioned at four angles of 0$^{\circ}$, 22.5$^{\circ}$, 45$^{\circ}$, and 67.5$^{\circ}$ per sequence.
Broad-band polarimetry was obtained with the $B$, $V$, $R$ and $I$ filters, available in both telescopes.
Spectropolarimetric observations with FORS2 were obtained using the 300V grism (with a wavelength coverage from $\sim$3300 to 9300 \AA), and with or without the order separating GG435 filter. The longpass filter GG435 has a cut-off at $\sim 4350$~\AA, and is used to prevent second-order contamination, which can be significant in the case of very blue objects\cite{Patat2010}, but is in most cases negligible. 
A detailed observing log is presented in Table~\ref{tab:log}.

The PMOS observations have been bias subtracted and reduced using standard IRAF/pyRAF procedures\cite{Cikota2019}. 
The ordinary and extraordinary beams of polarized spectra were extracted using the IRAF task \textsc{apextract.apall}, and wavelength calibrated using He-Ne-Ar arc lamp exposures. The typical RMS accuracy of the wavelength calibration is $\sim$0.3 \AA.
In addition to the polarization spectra, flux spectra were derived by summing the ordinary and extraordinary beams. The flux was calibrated using a generic sensitivity curve derived from flux standard stars observations observed in the PMOS mode. 
The noise in the polarization spectra has been reduced by applying a wavelet decomposition on the individual flux spectra of the ordinary and ordinary beams\cite{Cikota2019}.
We compared the denoised polarization spectra with the original and made sure that the method does not produce any systematic errors.
The broad-band polarimetry data from the VLT and the NOT were also reduced in a standard manner\cite{Leloudas2015,Leloudas2017} and fluxes were measured through aperture photometry.

Linear polarization can be described by the Stokes parameters $I$, $Q$ and $U$. The normalised Stokes parameters $q = Q/I$ and $u = U/I$ and their errors were calculated through the normalised flux differences\cite{Patat2006}. 
For the FORS2 data we corrected
for the retardance chromatism of the super-achromatic HWP using the wavelength-dependent retardance offset tabulated in the FORS2 User Manual\cite{FORS2manual}.
The polarization degree, $P=\sqrt{(q^2 + u^2)}$, is always overestimated  in the presence of noise, creating a bias known as the polarization bias. 
We adopt a correction $P_0 = (P - \sigma_P^2) \times h(P - \sigma_P)$, where $P_0$ is the bias-corrected polarization, $\sigma_P$ is the polarization uncertainty and $h$ is the Heaviside function\cite{Wang1997,Cikota2019}.

Our methodology and accuracy, as well as instrument stability, was cross-checked and verified by observations of standard stars (high polarization and zero polarization standards) obtained with the two telescopes in 2018 and 2019.

\subsection{NOT archival observations}

Two NOT broad-band polarimetry epochs of AT~2019dsg  were already presented in [Ref.\cite{Lee2020}]. The authors measured a polarization level of ($9.2 \pm 2.7$)\% on 2019-05-17, decreasing to $<$2.7\% on 2019-06-20, arguing for a rapid decrease in the polarization. While we agree with the qualitative conclusions of [Ref.\cite{Lee2020}], by re-analyzing this data  we cannot reproduce their quantitative results (Table~\ref{tab:quP}). 
The origin of this discrepancy is not clear. 
The ordinary and extraordinary beams in ALFOSC polarimetric data often have different PSF and ellipticities, depending probably on the observing conditions and the quality of the focus, which can result in erroneous polarization estimates if different fractions of light are included in aperture photometry for the two beams, especially at the low S/N regime. 
Our previous experience with the instrument, e.g. [Ref.\cite{Leloudas2017}], has shown that optimum results are obtained with apertures of 2 -- 2.5 times the FWHM of the ordinary beam PSF. 
The critical dataset from 2019-05-17 (S/N = 131) indeed suffers from this problem and by varying the aperture we do see a dependence of the result on the radius adopted.
However, we do get stable results (within one sigma) for our reference aperture interval and we never get values close to those of [Ref.\cite{Lee2020}], but values between 2--3\%.
Furthermore, the polarisation uncertainty we obtain is around $\sim$0.5\%, in line with the expectation from the S/N ratio\cite{Patat2006}.
Our NOT analysis methods are validated on our best dataset obtained on 2019-07-03. Excellent conditions (S/N $=$ 550 for the TDE) allow us to extract measurements for 7 additional stars in the field of view (including the one comparison star used in [Ref.\cite{Lee2020}]) and obtain a statistical estimate of the Galactic ISP, which is consistent with the one obtained with the superior VLT data (see below).

\subsection{Determination and removal of the interstellar polarization}

The ISP has been estimated in different ways. Our reference method was to estimate $q_{ISP}$ and $u_{ISP}$ as the weighted average of the Stokes parameters of field stars in the immediate vicinity of the TDE by using the imaging linear polarimetry data obtained at the VLT in different filters. This method was verified by checking catalogued polarization for nearby stars in the Heiles catalog\cite{Heiles2000}, by taking into consideration the nature of the TDE host galaxy, and by using the polarization spectrum of the TDE itself.    

The most challenging case is certainly AT 2018dyb, which, at Galactic latitude $b = - 7^{\circ}$, has a significant ISP contribution\cite{Holoien2020}.
Our methodology and results are presented below for each TDE. 

\textit{AT 2018dyb}. Broad-band imaging polarimetry has been obtained in the $B$, $V$ (twice) and $R$ filters. It is possible to measure the polarization for 15 -- 20 stars in the field of AT 2018dyb at projected distances between 0.8$^\prime$ -- 4.4$^\prime$ from the TDE and at high signal to noise (typically $> 700$). The field star polarization needs to be corrected for instrumental effects as FORS2 displays a well-known radial polarization pattern away from the optical axis of the instrument\cite{Patat2006,GonzalezGaitan2020} (we note that this correction does not affect the TDE itself which is placed in the centre of the field of view). After the instrumental polarization correction, we obtain the weighted average of the Stokes parameters as our best estimate for the Galactic ISP towards the TDE in each band. 
Subsequently we fit a Serkowski law\cite{Serkowski1975},  in order to determine the exact dependence of the polarization degree with wavelength. The formula used is $p(\lambda)/p_{\mathrm{max}} = \exp{[-K \ln^2{(\lambda_{\mathrm{max}}/\lambda)}]}$, where $p_{\mathrm{max}}$ is the maximum polarization at wavelength $\lambda_{\mathrm{max}}$ and the parameter $K$ can either be fit, considered a constant (typically $K = 1.15$) [Ref. \cite{Serkowski1975}], or related linearly to $\lambda_{\mathrm{max}}$, as $K = (-0.10 \pm 0.05) + (1.86 \pm 0.09)\lambda_{\mathrm{max}}$ [Ref. \cite{Whittet1992}].
In the case of AT 2018dyb, since we only have 3 wavelength points, we adopt the latter approach fitting all field stars simultaneously, and we obtain $p_{\mathrm{max}} = 1.08\%$ and $\lambda_{\mathrm{max}} = 6388$ \AA . Assuming that the polarization angle is constant (and equal to the average polarization angle 47.9$^{\circ}$ for $B$, $V$ and $R$), it is possible to obtain the final $q_{ISP}(\lambda)$ and $u_{ISP}(\lambda)$. These values are then vectorially subtracted from the Stokes spectra $q(\lambda)$ and $u(\lambda)$ to obtain polarization spectra corrected for the ISP. An independent check can be performed by checking polarized stars in the Heiles catalog. These are brighter stars (typically 10 mag) but the disadvantage is that their angular distance from the TDE is larger:  3 stars are found within 1$^{\circ}$ and 14 stars within 3$^{\circ}$. Their polarization properties seem to agree very well with our data and, notably, their polarization angles seem to cluster around 50$^{\circ}$, fully consistent with the value determined above. In addition, the star distances provided in the Heiles catalog span 1 -- 10 kpc indicating that the foreground stars probe the entire Milky Way sightline. There is no obvious bias of the polarization angle with respect to the recorded distance, apparent magnitude and $E(B-V)$. We are therefore confident that our ISP estimate accurately probes the Galactic contribution.  

This method has limitations as it neglects any ISP at the TDE host galaxy. The host galaxy of AT 2018dyb, however, is an old passive galaxy\cite{Leloudas2019,Holoien2020}, with no evidence of ongoing star formation, and with a limited probability of having any dust. Using the host galaxy spectrum we measure the strength of the 4000 \AA\ break and we find $D_{4000} > 1.53$. We therefore assume negligible reddening and ISP at the host.

We additionally note that the shape of the uncorrected polarization spectrum at day $+50$ is reminiscent of a Serkowski law, although the polarization angle varies considerably with wavelength (Extended Data Figures~\ref{fig:flux_pola_spec_orig} and \ref{fig:resultsall_orig}, panel b). 
We therefore tried another (limiting) case where the polarization at this phase is \textit{entirely} due to ISP 
and the intrinsic polarization of the TDE is equal to zero. 
By fitting the polarization spectrum we obtain 
$p_{\mathrm{max}} = 0.95\%$ and $\lambda_{\mathrm{max}} = 5502$ \AA. 
By assuming the average polarization angle, 
it is possible to determine $q_{ISP}(\lambda)$ and $u_{ISP}(\lambda)$ and remove this from the first epoch. 
This alternative solution results in the polarization spectra seen in Extended Data Figure~\ref{fig:flux_pola_spec_altISP}.
We have verified that this alternative solution does not alter any of the main conclusions of our paper (wavelength independence of the continuum polarisation, time evolution, etc). What is different is the overall polarization level, which becomes lower (about 1.2\% at $-17$ days), and the fact that the system appears axially symmetric already at $-17$ days (the dominant axis crosses the origin of the Stokes plane).
However, we do not favor this solution. As explained, polarization due exclusively to ISP, should have a wavelength-independent polarization angle (at least when due to a single dust cloud) and this is not what is observed for AT~2018dyb at $+50$ days.

\textit{AT 2019dsg}. Broad-band imaging polarimetry has been obtained in the $B$, $V$, $R$ and $I$ filters. 
The number of useful field stars ranged from 11 in the $B$ band to 23 in the $I$ band, at projected distances $<4.5^\prime$. Similar to AT 2018dyb, we applied an instrumental correction\cite{GonzalezGaitan2020} and derived the ISP in each band as the weighted average of the Stokes parameters of the field stars. Subsequently, we fit a Serkowski law to the $BVRI$ data and obtained  $p_{\mathrm{max}} = 0.16\%$ at $\lambda_{\mathrm{max}} = 3000$ \AA. We note that this  $\lambda_{\mathrm{max}}$ is at the blue limit of our allowed fit region as ISP  does not typically peak in bluer wavelengths\cite{Whittet1992}. 
The resulting ISP is generally small and other  fit choices do not significantly alter the ISP estimate at a level that is important for the results of this paper. The average polarization angle derived from the field stars is $-23.8^{\circ}$. By checking the Heiles catalog, we find 3 stars within 5$^{\circ}$ (only 1 within 3$^{\circ}$) and their polarization values are either consistent with zero, or at the level of $0.3 \pm 0.1\%$, confirming our more detailed estimate. Although AT 2019dsg is not found in a quiescent galaxy, we assume the host ISP to be negligible since there is no significant evidence for the opposite: the early TDE polarization spectrum shows minimal polarization, consistent with the Galactic ISP, at wavelengths $>7000$ \AA\ where the host completely dominates at all times ($a(\lambda) > 70\%$; Extended Data Figure~~\ref{fig:HostContrib}) . Furthermore, during the late broad-band observations, the location of the TDE on the $q$ -- $u$ plane is fully consistent with those of the foreground field stars. This would require the intrinsic TDE polarization and the host galaxy ISP to cancel out, which we think is unlikely.

\textit{AT 2019azh}. Unfortunately, no broad-band  linear polarimetry was obtained for AT 2019azh. It is therefore not possible to get an estimate of the ISP through field stars. There is evidence that any ISP is minimal: all nearby stars in the Heiles Catalog (only 3 stars are found within 5$^\circ$) have polarization consistent with zero. The Galactic extinction towards this direction is only  $E(B-V) = 0.039$ mag, placing an upper limit of $0.35\%$ to the Galactic ISP (using $P < 9 \cdot E(B-V)$)\cite{Serkowski1975}. In addition, the TDE polarization spectrum shows little polarization in the redder wavelengths where the light is dominated by the host galaxy. Finally, the host is an E+A galaxy\cite{Hinkle2020}, similar to many TDE hosts\cite{Arcavi2014,French2016}, with no evidence for current star formation, and the X-ray derived column density $N_H$ does not suggest any enhanced extinction but is consistent with the Galactic value along the line of sight, similar to most TDEs\cite{Auchettl2017}.
We therefore do not consider any ISP contribution for AT 2019azh  
as our reference hypothesis.

As an alternative, we attempted an ISP solution where the H$\alpha$ line is completely depolarized at the core. This is an assumption often made in the literature for supernova explosions. This solution is shown in Extended Data Figure~\ref{fig:flux_pola_spec_altISP} and we have verified that it only has minor quantitative effects and does not alter the qualittative conclusions of this study. We do not favor a solution where the H$\alpha$ line is completely depolarized already at $+22$ days as this line becomes stronger with time\cite{Hinkle2020}. Irrespective, the effects of ISP for AT 2019azh seem negligible.  

\subsection{Estimation of light dilution from the host galaxy} In order to estimate the polarization intrinsic to the transient we need to correct for the effect of light dilution from the host galaxy, as the integrated stellar light is not expected to be polarized. The host galaxy contributes significantly at late times and at  redder wavelengths, as TDEs are generally blue while their hosts are typically red. By defining the host contribution as the ratio $\alpha(\lambda) = I_{host}(\lambda)/I_{tot}(\lambda)$, where the total flux is $I_{tot}(\lambda) = I_{host}(\lambda) + I_{TDE}(\lambda)$, it is possible to show that for the Stokes parameters we have $q_{TDE}(\lambda) = q_{tot}(\lambda)/(1 - \alpha(\lambda))$ with a similar relation for $u(\lambda)$. The dilution correction therefore assumes $q_{host} = 0$ and it has been extensively discussed in polarization studies of AGN\cite{MillerGoodrich,Marin2018}.

In order to determine the host contribution $\alpha(\lambda)$ we simply divide a spectrum of the host with the VLT spectrum including the TDE (after proper absolute flux calibration).
In the case of AT 2019azh we used the archival SDSS host spectrum. 
In the case of AT 2018dyb we used a spectrum at $+$544 days obtained after the TDE had faded\cite{charalampopoulos2022}. The wavelength range of this spectrum, however, is smaller than the VLT/FORS2 spectrum, reducing the useful wavelength range after this correction was applied.
For AT 2019dsg, we used a spectrum obtained with the NOT at $+730$ days. 
In all cases, we have scaled the spectra with appropriate multi-band photometry before computing the host contribution. 
The resulting $\alpha(\lambda)$ are shown in Extended Data Figure \ref{fig:HostContrib}, where it can be appreciated that this  correction is important:
the host contribution ranges from 10--20\% in the blue and up to $>$70\% in the red for some TDEs and phases, showing a significant depression at the location of strong emission lines.
We caution that in the case of AT 2019dsg, our host galaxy subtraction was imperfect at the region of the H$\alpha$ line due to the presence of both the narrow component and the telluric B-band.

Although we have done our best to apply the ISP and host dilution corrections, these are indeed corrections that are challenging in practice and they can  alter the shape of the TDE polarization spectrum (for instance, without the host correction the polarization spectrum has a strong wavelength dependence).
For this reason, we also present our uncorrected data to our readers in Extended Data Figures~\ref{fig:flux_pola_spec_orig}, \ref{fig:resultsall_orig}, \ref{fig:Ha_He_orig} and  \ref{fig:Stokes_norot_original}. 

\subsection{The TDE disk model}
In most tidal disruption events we expect that the debris fallback rate largely exceeds the Eddington accretion rate $\dot{M}_{\rm Edd}$. As a result, geometrically thick disks can be formed due to the large radiation pressure.
As analytical models of super-Eddington disks are lacking, we adopt a TDE super-Eddington disk  simulated using the \textsc{HARMRAD} code\cite{Mckinney14} from a previous study\cite{Dai18}.  \textsc{HARMRAD} is a state-of-the-art 3D general relativistic radiation magnetohydrodynamic (GRRMHD) code capturing electon scattering, Comptonization and basic emission and absorption physics. The simulated disk has the following parameters: black hole mass $M_{\rm BH}=5\times10^6M_\odot$, black hole spin $a=0.8$, accretion rate $\dot{M}_{\rm acc}\approx 15 \dot{M}_{\rm Edd}$, outflow rate $\dot{M}_{\rm wind}\approx 10 \dot{M}_{\rm Edd}$, disk mass $m_{\rm disk} \sim 0.2 M_\odot$ and the disk specific angular momentum consistent with that of the bound debris. The disk is circular and aligned with the black hole spin.

The different TDE disks we use to do polarization calculations throughout this work are all based on 
this reference disk model with some modifications.
For the original model which has relatively extended disk and wind profiles (called ``extended disk model'' hereafter), we only vary the magnitude of the density, so the disk mass changes as $m_{\rm disk} \propto \rho$. The disk masses used to do polarization calculations are scaled to be $0.01-0.1 M_\odot$ as reasonable choices of TDE disk masses. The density of an extended disk with a total disk mass of $0.1 M_\odot$ is plotted Figure \ref{fig:models_geometry}(a). Its electron scattering photosphere with $\tau_{\rm ES} =1$ (integrated radially inwards from faraway) is also plotted over the density. Alternatively, we shrink the disk size by ten times while keeping the same structure of the density (which are called ``compact disk model'' hereafter). Therefore, the density of a compact disk is much higher compared to the density of an extended disk of the same mass. Since optical depth $\tau \propto \rho R$, the compact disk has a larger optical depth than the extended disk. The density and scattering photosphere of a compact disk with a mass of $0.1 M_\odot$ are shown in Figure \ref{fig:models_geometry}(b).

Recent numerical simulations\cite{Ohsuga2011, Jiang2014} universally show that super-Eddington disks launch winds with density increasing with inclination. Therefore, a optically thin ``funnel'' is naturally produced in the polar region, and outside the funnel the optically and geometrically thick disk and wind reprocess the X-ray photons produced in the inner disk region. This can be used to explain the origin of UV/optical emissions observed in TDEs\cite{Strubbe2009, LodatoRossi}. In particular, it has been shown\cite{Dai18} that for the simulated disk we use, strong X-rays can only leak out when the observer is looking down the funnel, while UV/optical emission dominates when the disk is viewed at large inclination.

\subsection{Polarization simulation}

All the polarization modelling presented in this work was carried out using \textsc{possis}\cite{BullaPOSSIS}, a 3-D Monte Carlo radiative transfer code that has been used in the past to predict linear polarimetry of both supernovae\cite{Bulla2015,Inserra2016} and kilonovae\cite{BullaKN,Bulla2021}. The code uses input opacities to determine what fraction of the radiation produced can escape the system and is flexible enough to accommodate arbitrary 3-D geometries. The radiation is represented by indivisible Monte Carlo photon packets that are created with initial location sampled throughout the envelope according to the specific density distribution within the model. This assumption is chosen to simulate the expected reprocessing of X-rays to optical photons in TDEs. Monte Carlo photons are then followed as they propagate through the expanding medium and interact with matter via electron (Thomson) scattering, bound-bound line transitions or bound-free and free-free continuum absorptions. We focus on the continuum polarization and neglect bound-bound opacities.

While electron scattering is responsible for polarizing radiation, an absorption component from bound-free and free-free transitions can depolarize the radiation and possibly introduce a wavelength dependence in the overall polarization signal. 
Since we do not see a significant wavelength dependence in the continuum polarization of our three TDEs, we neglect absorption opacities in our modelling and assume a pure electron-scattering case for our reference models (see the last section for a more detailed justification on this). Assuming solar composition (hydrogen mass fraction $X=0.7$) as in [Ref.\cite{Dai18}], 
we set the electron-scattering opacity to $\kappa_\mathrm{es}= 0.2 \times(1+X)=0.34$\,cm$^2$\,g$^{-1}$. When a photon packet is scattered by an electron, it acquires linear polarization in a direction perpendicular to the scattering plane. The linear polarization is described by the Stokes parameters $I$, $Q$ and $U$, which are properly updated after each interaction with an electron\cite{Bulla2015,BullaPOSSIS}. In place of using the standard angular binning of the escaping photons, here we adopt the ``virtual-packet" approach described in [Ref.\cite{Bulla2015}] to extract polarization levels for different viewing angles. Synthetic observables predicted with this technique have been shown\cite{Bulla2015} to be more accurate and less affected by Monte Carlo noise, thus reducing the computational cost of each simulation. All the simulations presented in this work have been carried out for 11 viewing angles in the $xz$ plane, equally spaced in cosine between a polar ($\cos\theta=1$, face-on) and an equatorial ($\cos\theta=0$, edge-on) orientation ($\Delta\cos\theta=0.1$). The final Stokes parameters $I$, $Q$ and $U$ for each observer are computed by summing contributions from each photon and the normalized $q$ and $u$ are calculated as $q=Q/I$ and $u=U/I$. The axial symmetry of the disk
models is such that the contributions to the $U$ Stokes parameter are cancelled and the final $u$ signal is consistent with zero. The polarization signal is thus carried by $q$, while $u$ is used as a proxy for the Monte Carlo noise. We note that the models are assumed to be static and effects that could be introduced by disk rotation / winds are not considered.

Extended Data Figure~\ref{fig:Qmodels} shows the polarization degree $q$ as a function of viewing angle for the extended disk model with five different disk masses: $m_\mathrm{disk}=0.01$, $0.0125$, $0.025$, $0.05$, $0.1\,M_\odot$ around a black hole of $M_{\rm BH}=5\times10^6\,M_\odot$.
In all simulations, $Q=0$ for a polar orientation ($\cos\theta=1$, face-on inclination) due to the axial symmetry of the model, while non-zero polarization signals are found for different orientations. When the inclination angle or the disk mass increases, the seed photons need to travel through denser gas flows, accentuating the multiple electron scatterings. This generally increases the polarization of the escaped photons. 
For example, for an equatorial viewing angle ($\cos\theta=0$), we see a general increase in the absolute value of polarization, e.g. from $\sim2.5\%$ to $\sim6\%$ as the disk mass goes from $m_\mathrm{disk}=0.01\,M_\odot$ to $m_\mathrm{disk}\gtrsim 0.05\,M_\odot$. 
As the TDE disk mass likely correlates with the debris fallback rate, one can expect that the disk mass decreases at late time in TDEs, which leads to a decrease in polarization as well.  This predicted trend is consistent with the polarization evolution observed from the TDEs.

Polarization $q$ levels are preferentially negative. This can be understood by inspecting the density distribution for the case of $m_\mathrm{disk}=0.1\,M_\odot$ shown in Figure~\ref{fig:models_geometry}. The simulated super-Eddington disk is characterized by high-density puffed-up disks around the equator and winds at lower densities closer to the pole. As a result, photons preferably leak from the high-opacity region (disk) to the low-opacity region (wind) due to radiation pressure. Also, photons travelling through the lobes experience multiple scattering with electrons, effectively causing a loss of information on directionality and thus destroying the polarization signal\cite{Kasen2003}. On the contrary, photons travelling through the funnel are less affected by multiple scattering and are scattered towards the observer with an electric field oscillating in the horizontal direction, i.e. with a negative $Q$. The combination of $Q<0$ contributions from the funnel and little polarizing contributions from the lobes biases the overall polarization signal to negative $Q$ values.

\subsection{Alternative explanations for the optical polarization} This section examines origins of polarization alternative to electron scattering, including synchrotron, dust and absorption.

\textit{Synchrotron.} The possibility that the optical polarization in AGNs could be due to synchrotron was considered initially\cite{Walker1968,MillerAntonucci1983} but rejected early, already in the seminal paper that led to the AGN unification model\cite{AntonucciMiller1985}. As a small number of optical TDEs have been detected in the radio\cite{Alexander2020} and this emission has in some cases been attributed to synchrotron radiation\cite{Stein2020}, we provide evidence that any synchrotron contribution in the optical would be minimal. AT~2019dsg is indeed the strongest radio emitter among optical TDEs. Using data from [Ref. \cite{Stein2020}] we observe that $F_{\nu,\mathrm{opt}} / F_{\nu,\mathrm{radio}} \sim 0.4$ at 42 days decreasing to 0.04 at 178 days (where $F_{\nu,\mathrm{opt}}$ refers to the $g$ band and $F_{\nu,\mathrm{radio}}$ to 15 GHz). Extrapolating the synchrotron power-law fit to the optical range we find that the contribution of any synchrotron component in the optical is negligible ($< 0.1$\%). Even considering the theoretical limiting case where this component would be very strongly polarized ($\sim70$\%), this would have no observable effect in the optical polarization. AT~2019azh has also been detected in the radio, producing even a late-time ($>200$ days) flare\cite{sfaradi2019azh}, but especially during  our spectropolarimetric observations the radio component is much weaker than in AT~2019dsg ($F_{\nu,\mathrm{opt}} / F_{\nu,\mathrm{radio}} \sim 10$), pointing to the same conclusion. AT~2018dyb, despite its proximity, was not detected in the radio\cite{Holoien2020,Alexander2020}. We conclude that polarization due to a synchrotron component is not important for optical TDEs. 

The situation will be different for relativistic TDEs\cite{Zauderer2011,Cenko2012} where synchrotron might indeed be the dominant source of polarization in the optical/NIR regime\cite{Wiersema2012,Wiersema2020}.

\textit{Dust.} Scattering by dust is also an important source of polarization\cite{Serkowski1975} and this was already discussed in the section investigating the ISP. Here we focus on insights from MIR observations and dust reverberation \cite{vanvelzenTDEreverb,Jiang2021}.
Supermassive black holes in AGN are surrounded by a dusty torus and TDEs occurring in AGN can produce a strong signal in MIR when the TDE light echoes on this pre-existing dust
\cite{Jiang2017}. The potential existence of such a torus in quiescent galaxies is an unsolved matter. A systematic search\cite{Jiang2021} with NEOWISE included all optical TDEs until 2018 and concluded that the dust covering factor in the sub-pc scale is much smaller ($f_c \sim 0.01$) in TDEs found in quiescent galaxies than in star-forming galaxies\cite{Wang2022} and AGNs. AT~2018dyb was included in this study and was detected in both the W1 and W2 filters around peak (the NEOWISE cadence is 6 months). The authors argue that the MIR component is above the Rayleigh-Jeans tail of the optical black-body ($T_{\rm dust}\sim 1450$~K) but they compute a dust covering factor of barely 0.3\%, corroborating our conclusion that dust is unlikely to contribute significantly in the polarization of AT~2018dyb. We have ourselves constructed the NEOWISE light curves of AT~2019azh and AT~2019dsg. AT~2019azh is very similar to AT~2018dyb, showing a weak detection, likely due to the proximity of both events, and yielding a dust covering factor of $\sim$0.5\%. AT~2019dsg, found in a star-forming galaxy or possible AGN, is very different and shows a strong and long-lasting MIR echo peaking $\sim$170 days after the optical maximum. Nevertheless, the dust covering factor is $\sim$3\%, which is again rather low. Furthermore, the fact that the optical polarization degree decreases rapidly with time (Figure~\ref{fig:timevol}), while the MIR emission increases with time (and at much longer timescales), convincingly argues for scattering by dust not being responsible for the optical polarization. Finally, there is the issue of the apparent wavelength independence of the optical polarization. Polarization by dust should be wavelength-dependent, although this depends strongly on the size and distribution of the dust grains and the observed wavelength range is rather short to exclude different configurations, which would likely require observations in the UV\cite{GoosmannGaskell2007,Zubko2000}.   

\textit{Depolarization by bound-free and free-free transitions.} We here investigate whether our assumption to ignore an absorption component from bound-free and free-free transitions in our reference models can be justified based on the apparent wavelength independence of the optical polarization. 
If the absorption opacity
($\kappa_{\rm abs}=\kappa_{\rm bf}+\kappa_{\rm ff}$)
is comparable to that from electron scattering ($\kappa_{\rm es}$), it is possible that a wavelength dependence can be introduced in the overall polarization signal. 
To check this we performed tests with three opacity models. 
Model 0 (ES) includes only electron scattering in the polarization calculation. For model 1 and 2 we include not only electron scattering but also absorption, and the absorption opacity is assumed to increase with wavelength following the same power-law function as the free-free absorption opacity, as illustrated in Extended Data Figure~\ref{fig:wavedep_a}(a). Model 1 (ES+Abs1) has $\kappa_{\rm abs}/\kappa_{\rm es}<0.1$ and model 2 (ES+Abs2) has $\kappa_{\rm abs}/\kappa_{\rm es} \sim 1$ at 7000~\AA. 
This is approximately 3 orders of magnitude above what was computed by [Ref. \cite{Roth16}] but we include this in our investigation as a limiting case.
For all three models, we run the polarization calculations with a full set of parameters ($m_{\rm disk}=0.01, 0.02, 0.03, ..., 0.1 M_\odot$ and $\cos{\theta_{\rm obs}}=0, 0.1, 0.2, ..., 1.0$).  We then fit the polarization of the continuum part of AT2018dyb (at $-$17 days) 
to the models, and a number of fits for the three opacity scenarios are shown in Figure~\ref{fig:wavedep_a}(b). 
The best fit is given by Model 0 (ES) with a reduced chi-square of $\chi^2_\nu = 0.7$ and $m_{\rm disk} = 0.02 M_\odot$ seen $\sim55^{\circ}$ from the pole. Models 1 (ES+Abs1) and 2 (ES+Abs2) yield worse fits and a wavelength dependence for Model 2, using the same parameters. 
It is, however, possible to obtain good fits for alternative regions of the parameter space (disk mass and inclination angle) that show small wavelength dependence, especially for Model 1. As the observed wavelength region is rather short, and the model parameter space is big, we cannot exclude solutions with large $\kappa_{\rm abs}$. It is however, most likely that that electron scattering opacity dominates in the TDE photosphere. 

\end{methods}


\clearpage
 
\begin{addendum}

\item[Correspondence] Correspondence
should be addressed to Giorgos Leloudas~(email: giorgos@space.dtu.dk) and Lixin Dai~(email: lixindai@hku.hk).

\item[Data availability statement] All raw data are publicly available through the ESO and NOT archives. Reduced data are available from the first author on reasonable request.

\item[Code availability statement] The radiative transfer code POSSIS used in this work is not publicly available. Results presented in this work are available from Mattia Bulla upon reasonable request.

\item 
We thank Nando Patat and Santiago Gonz{\'a}lez-Gait{\'a}n for discussions concerning instrumental polarization corrections.
We acknowledge use of routines from the FUSS code (\url{https://github.com/HeloiseS/FUSS}) by Heloise Stevance.
GL, PC and DBM were supported by a research grant (19054) from VILLUM FONDEN. 
MB acknowledges support from the Swedish Research Council (Reg. no. 2020-03330).
LD and LT acknowledge support from the Hong Kong RGC (GRF grant HKU27305119 and HKU 17304821) and the NSFC Excellent Young Scientists Fund (HKU 12122309).
Lawrence Livermore National Laboratory is operated by Lawrence Livermore National Security, LLC, for the U.S. Department of Energy, National Nuclear Security Administration under Contract DE-AC52-07NA27344. 
IA is a CIFAR Azrieli Global Scholar in the Gravity and the Extreme Universe Program and acknowledges support from that program, from the European Research Council (ERC) under the European Union’s Horizon 2020 research and innovation program (grant agreement number 852097), from the Israel Science Foundation (grant number 2752/19), from the United States - Israel Binational Science Foundation (BSF), and from the Israeli Council for Higher Education Alon Fellowship.
Parts of this research were supported by the Australian Research Council Centre of Excellence for All Sky Astrophysics in 3 Dimensions (ASTRO 3D), through project number CE170100013.
DBM acknowledges support from ERC grant number 725246.
MN acknowledges funding from the European Research Council (ERC) under the European Union’s Horizon 2020 research and innovation programme (grant agreement No.~948381) and a Fellowship from the Alan Turing Institute.
Based on observations collected at the European Organisation for Astronomical Research in the Southern Hemisphere under ESO programmes 0102.D-0116(A) and 0103.D-0350(A).
Based on observations made with the Nordic Optical Telescope, owned in collaboration by the University of Turku and Aarhus University, and operated jointly by Aarhus University, the University of Turku and the University of Oslo, representing Denmark, Finland and Norway, the University of Iceland and Stockholm University at the Observatorio del Roque de los Muchachos, La Palma, Spain, of the Instituto de Astrofisica de Canarias.

\item[Author Contributions] 
GL initiated the project, was PI of the observing proposals, reduced all broad-band polarimetry, analysed the data and wrote most of the manuscript.
MB provided the radiative transfer models, and contributed with text and figures.
AC reduced all spectral polarimetry, helped with the data analysis and contributed text.
LD initiated the project with GL, provided the disk models and contributed with text. 
LLT contributed to the disk modeling and MCMC fitting.
JM helped with data analysis and interpretation. 
PC helped with the host correction and figure production.
NR contributed with theoretical predictions and interpretation. 
DBM helped with observation coordination and triggered the telescope.
All authors contributed in discussions at different stages of the project and provided comments on the manuscript.

\item[Competing Interests] The authors declare that they have no
competing financial interests.
This document was prepared as an account of work sponsored by an agency of the United States government. Neither the United States government nor Lawrence Livermore National Security, LLC, nor any of their employees makes any warranty, expressed or implied, or assumes any legal liability or responsibility for the accuracy, completeness, or usefulness of any information, apparatus, product, or process disclosed, or represents that its use would not infringe privately owned rights. Reference herein to any specific commercial product, process, or service by trade name, trademark, manufacturer, or otherwise does not necessarily constitute or imply its endorsement, recommendation, or favoring by the United States government or Lawrence Livermore National Security, LLC. The views and opinions of authors expressed herein do not necessarily state or reflect those of the United States government or Lawrence Livermore National Security, LLC, and shall not be used for advertising or product endorsement purposes.

\end{addendum}


\newpage

\begin{table}
    \centering
    \begin{tabular}{|c|c|c|c|c|c|c|c|}
    \hline
    \textbf{UT date} & \textbf{MJD}  & \textbf{Phase (d)}  & \textbf{Telescope} & \textbf{Grism / Filter} & \textbf{Exp. time (s)} & \textbf{Seeing ($''$)} & \textbf{S/N} \\
    \hline

    \multicolumn{8}{|c|}{\textbf{AT 2018dyb }}  \\    
    \hline
    2018-07-24 &  58323.0  &    $-$17.4 &   VLT &   300V, GG435    & 3x700 & 0.8--0.9   & 850 \\ 
    2018-09-18  &  58379.0 & +37.6 &   VLT &   $V$    & 2x40 & 0.6   & 1535 \\    
    2018-09-29  &  58390.0  & +48.4 &   VLT $^{1}$ &   300V    & 900 & 0.8--1.0   & 520 \\    
    2018-09-30  &  58391.0  & +49.4 &   VLT $^{1}$ &   300V    & 900 & 0.5   & 620 \\    

    2019-02-10  &   58524.3 & +180.3 &   VLT &   $B$, $V$, $R$    & 3x80, 3x50, 3x50 & 0.5--0.7   & 767 \\    

    \hline
    \multicolumn{8}{|c|}{\textbf{AT 2019azh}}  \\    
    \hline
    2019-04-08   &   58581.0  & +22.0 &   VLT &   300V    & 2x800 & 0.6-1.2   & 1020 \\ 

    \hline
    \multicolumn{8}{|c|}{\textbf{AT 2019dsg}}  \\    
    \hline
    2019-05-17  &   58620.2 & +16.3 &   NOT $^{2}$ &   $V$    & 100 &  1.5  & 132 \\    
    2019-05-28  &   58631.2 & +26.7 &   NOT $^{3}$ &   $V$    & 300 & -  & - \\    
    2019-06-02  &   58636.3  & +31.6 &   VLT &   300V    & 3x700 & 0.7--0.9   & 500 \\    
    2019-06-20  &   58654.1 & +48.5  &   NOT $^{2}$ &   $V$    & 200 &  0.9  & 194 \\    
    2019-07-03  &   58667.1 & +60.9  &   NOT &   $V$    & 400 &  1.2  & 550 \\   
    2019-07-13  &   58677.1 & +70.4 &   VLT &   $B$, $V$, $R$, $I$    & 225 -- 480 & 0.6--0.7   & 575--767 \\    
    \hline

    \end{tabular}
    \caption{A log of our observations. Spectropolarimetry epochs are those that include grism 300V. All other epochs are broad-band polarimetry. The exposure time is given per HWP. The S/N for spectropolarimetry refers to 5500 \AA\ and to a  25 \AA~bin. \textit{Notes}: $^{1}$ These two epochs have been co-added to increase S/N. The resulting S/N ratio is 820.
    $^{2}$ Data from [Ref.\cite{Lee2020}] (re-analyzed).
    $^{3}$ These data were obtained during twilight, are of low quality and give erroneous results. They were not used. }

    \label{tab:log}
\end{table}

\newpage

\begin{table}

    \centering
    \begin{tabular}{|c|c|c|c|c|c|c|c|}
    \hline
    Phase (d) & $q$ (\%)  & $u$ (\%)  & $\alpha(V)$ &  $q$ (\%) & $u$ (\%) & $P_0$ (\%) & $\sigma_P$ (\%) \\
    \hline
     & \multicolumn{2}{c|}{\textbf{fully reduced}} & \multicolumn{5}{c|}{\textbf{ISP+host corrected}} \\
    \hline
    
    \multicolumn{8}{|c|}{\textbf{AT 2018dyb }}  \\    
    \hline
    $-$17.4 $^*$ & $-$1.12 (0.01)  & 0.24 (0.01) & 0.34 (0.07) & $-$1.59 (0.17)  &  $-$1.35 (0.14)   & 2.07  & 0.16\\
    +37.6 & $-$0.80 (0.05)  & 0.87 (0.05)  & 0.37 (0.07) &   $-$1.16 (0.15)  &  $-$0.41 (0.09)  & 1.21  & 0.15\\
    +49.4 $^*$ & $-$0.62 (0.01)  & 0.72 (0.01) & 0.50 (0.11) &  $-$1.10 (0.24)  &  $-$0.82 (0.18) & 1.34  & 0.22\\
    +180.3 & $-$0.79 (0.09)  & 0.68 (0.09) & 0.94 (0.19) & $-$12.00 (38.03)  &   $-$7.50 (23.80)  & 0.00  & 34.63\\
    \hline
    
    \multicolumn{8}{|c|}{\textbf{AT 2019azh}}  \\    
    \hline
    +22.0 $^*$ & 0.02 (0.01)  & 0.45 (0.01) & 0.35 (0.05) & 0.03 (0.01)  & 0.69 (0.05)  & 0.68  & 0.05 \\
    \hline

     \multicolumn{8}{|c|}{\textbf{AT 2019dsg}}  \\    
    \hline   
    +16.3   &  2.72 (0.54) &  $-$0.39 (0.53) & 0.59 (0.02)& 6.41 (1.35)  &  $-$0.71 (1.30) &  6.17 & 1.35 \\
    +31.6 $^*$ & $-$0.15 (0.01)  &  0.29 (0.01) & 0.64 (0.02) & $-$0.67 (0.07) & 1.08 (0.11) &  1.26  & 0.10\\
    +48.5 & $-$0.64 (0.36)  &  0.13 (0.36) & 0.70 (0.03) & $-$2.43 (1.23) & 0.77 (1.21)  &  1.96 & 1.22\\
    +60.9 &  0.16 (0.13) &  0.00 (0.13) & 0.73 (0.02) &  0.26 (0.49) &  0.37 (0.49) &  0.00 & 0.49 \\
    +70.4 &  $-$0.01 (0.09) &  0.14 (0.09) & 0.75 (0.02) & $-$0.40 (0.37)  & 0.96 (0.39)  &  0.90 & 0.38 \\
    \hline

    \end{tabular}
    
    \caption{{Normalized Stokes parameters and bias-corrected polarization degrees for the three TDEs in the $V$ band. The host contribution $\alpha(V)$ at the $V$ band is also provided. Synthetic broad-band polarimetry has been computed at the epochs of spectral polarimetry (marked with an asterisk $^*$) by convolving with the filter function. We provide both the Stokes parameters as measured in the fully reduced spectra and after the ISP and the host correction. The polarization degree $P=\sqrt{(q^2 + u^2)}$ has been corrected for polarization bias using $P_0 = (P - \sigma_P^2) \times h(P - \sigma_P)$.}}
    \label{tab:quP}

\end{table}


\clearpage

\newpage

\newpage

\newpage

\newpage

\clearpage

\begin{addendum}

\item[]
%

\begin{SIfigure}
\centering
\includegraphics[width=\textwidth]{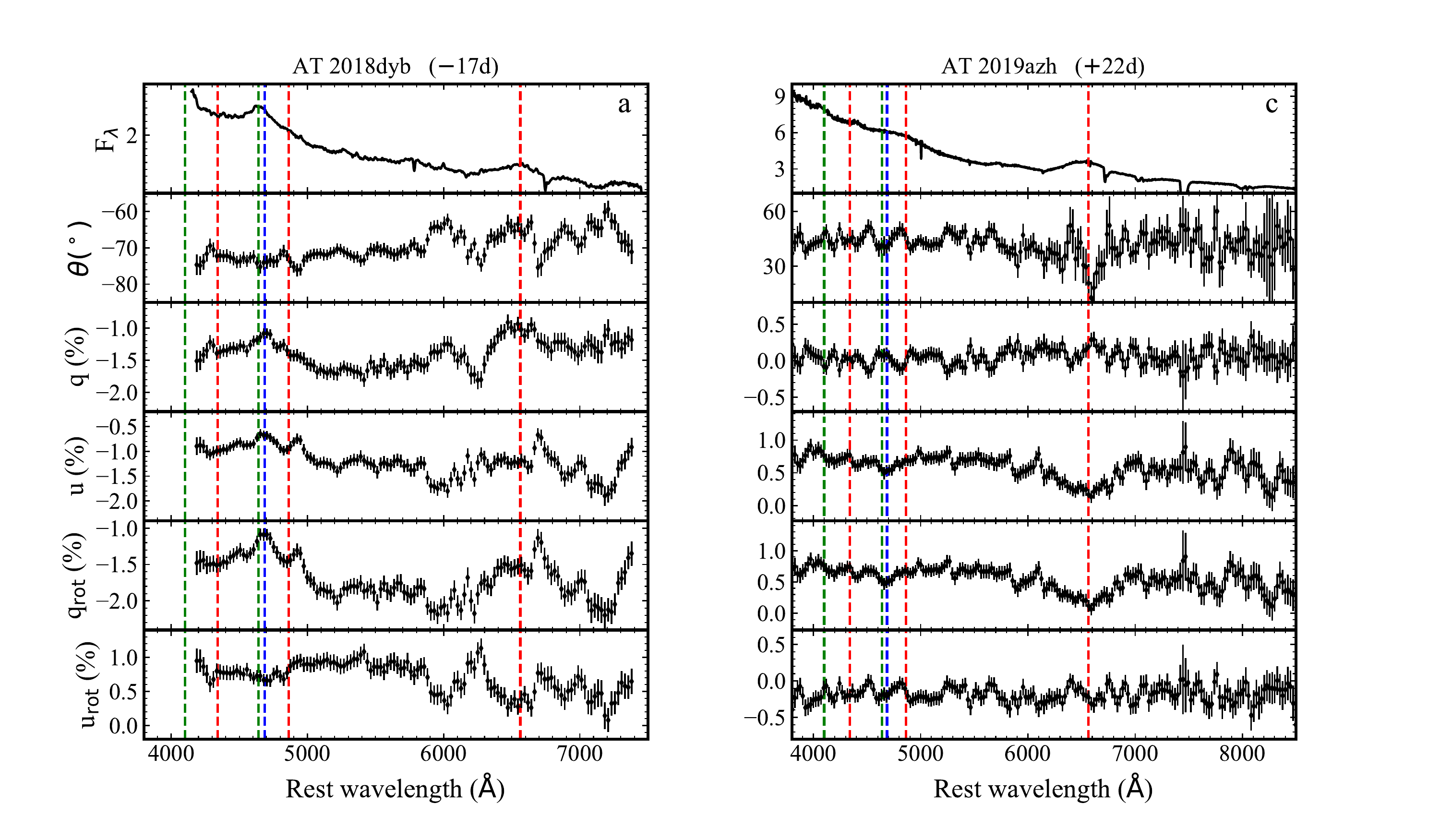}
\includegraphics[width=\textwidth]{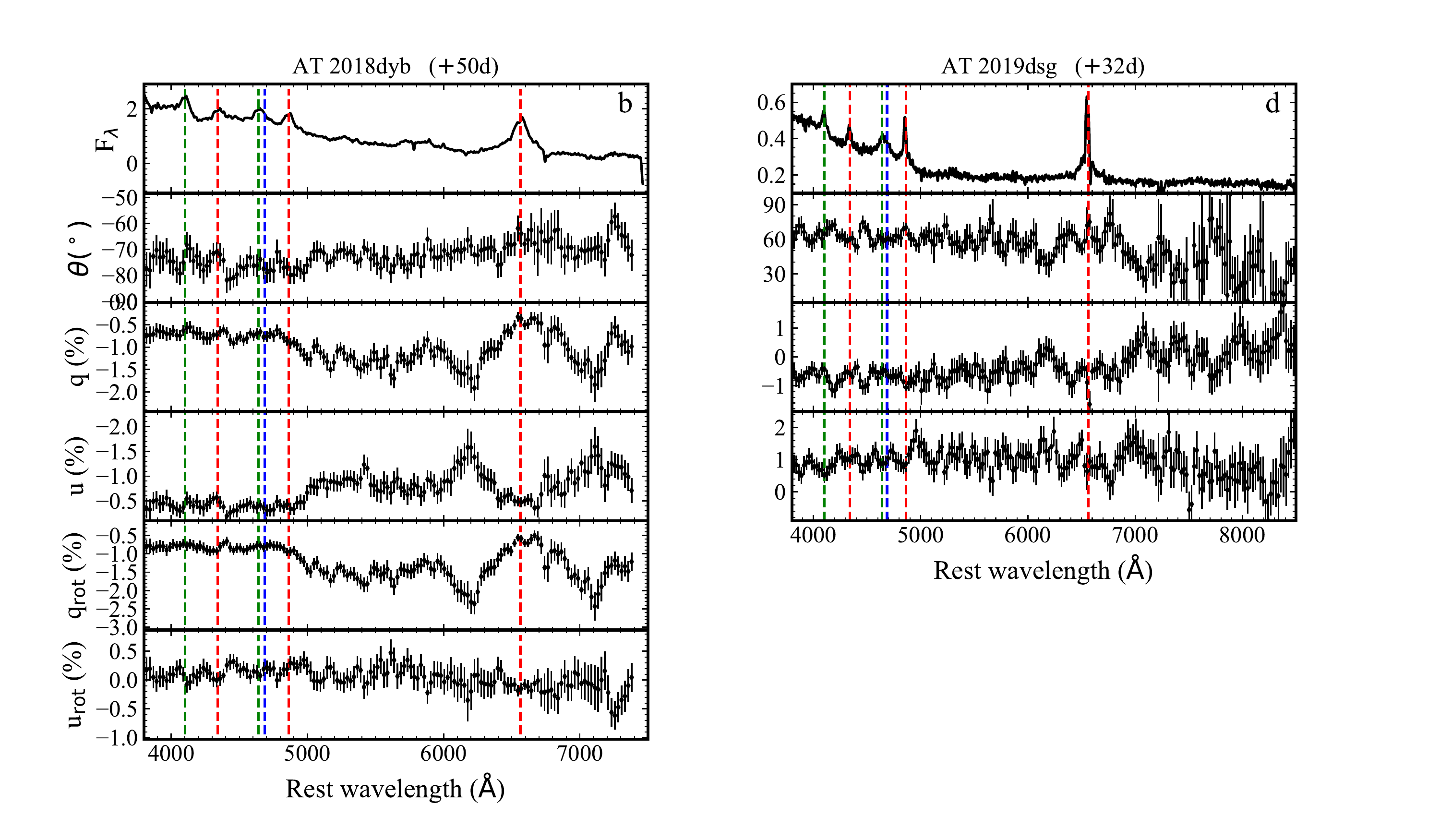}
\caption{\boldmath$\vert$\unboldmath \hspace{1em} \textbf{Polarization properties as a function of wavelength.} 
This figure is similar to Figure~\ref{fig:flux_pola_spec}.  The different panels below the flux spectra show the polarization angle $\theta$, the normalized Stokes parameters $q$, $u$, and the rotated Stokes parameters (from Extended Data Figure~\ref{fig:Stokes_rot_corrected}).
Dashed lines mark the wavelengths of major emission lines: red for Balmer lines, blue for He~II and green for N~III.
All error bars are 1$\sigma$ uncertainties and F$_\lambda$ is in units of erg s$^{-1}$ cm$^{-2}$ $\mathrm{\AA}^{-1}$.
All data is shown here after correcting for the ISP and for the host dilution.
\label{fig:resultsall_cor}
}
\end{SIfigure}

\newpage

\begin{SIfigure}
\centering
\includegraphics[width=16cm]{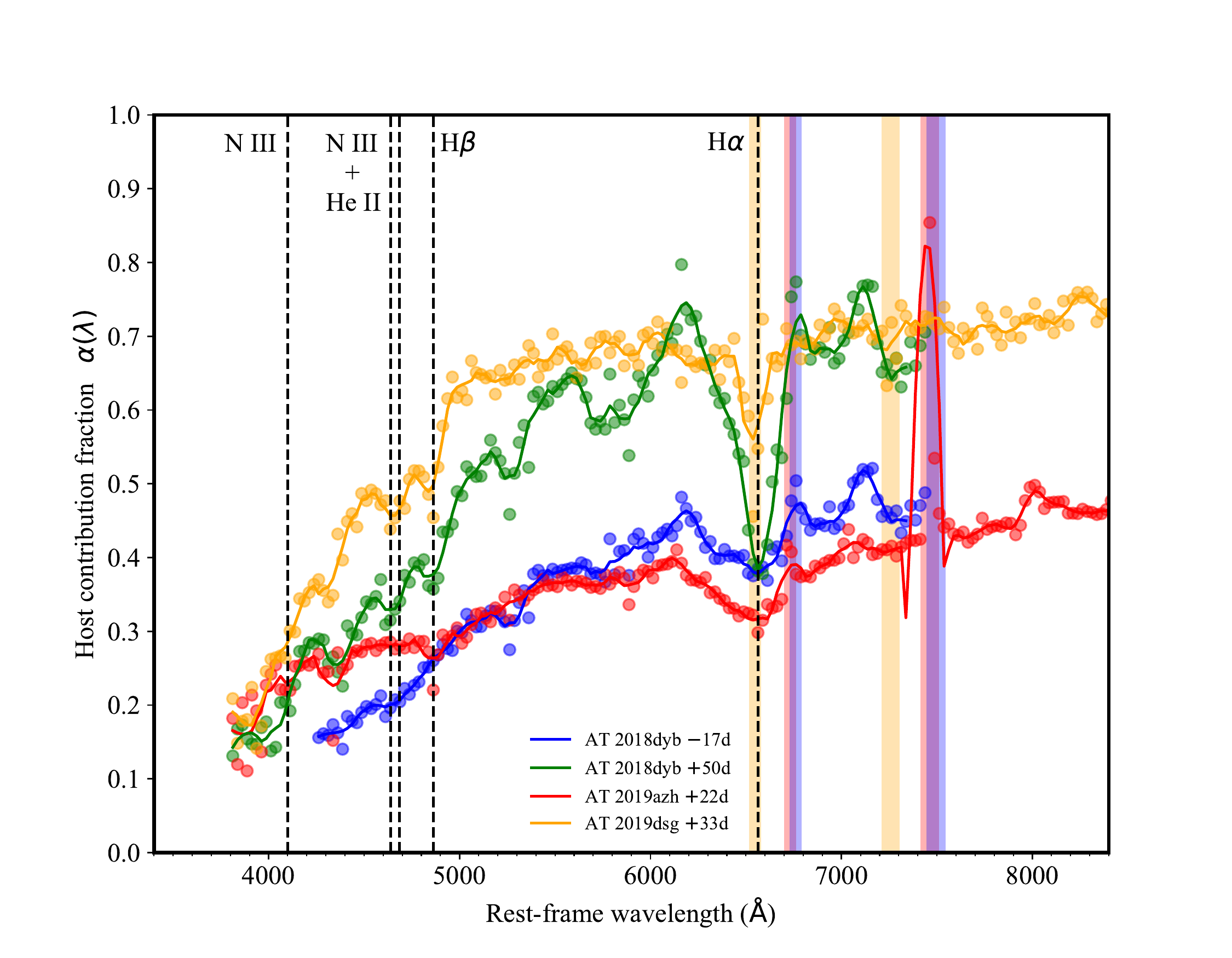}
\caption{\boldmath$\vert$\unboldmath \hspace{1em} \textbf{Host galaxy contamination at the time of the VLT spectropolarimetry.} 
The host contribution  is defined here as the flux ratio $\alpha(\lambda) = I_{host}(\lambda)/I_{tot}(\lambda)$ and is computed by dividing a spectrum of the host with the VLT spectrum including the TDE (after proper absolute flux calibration). Solid lines represent smoothed versions for each dataset, using a Savitzky-Golay filter. The location of prominent emission features is marked with vertical dashed lines and regions of significant telluric absorption are shown as shaded regions (their location differs for each TDE due to their different redshifts).  
\label{fig:HostContrib}
}
\end{SIfigure}

\newpage

\begin{SIfigure}
\centering
\includegraphics[width=\textwidth]{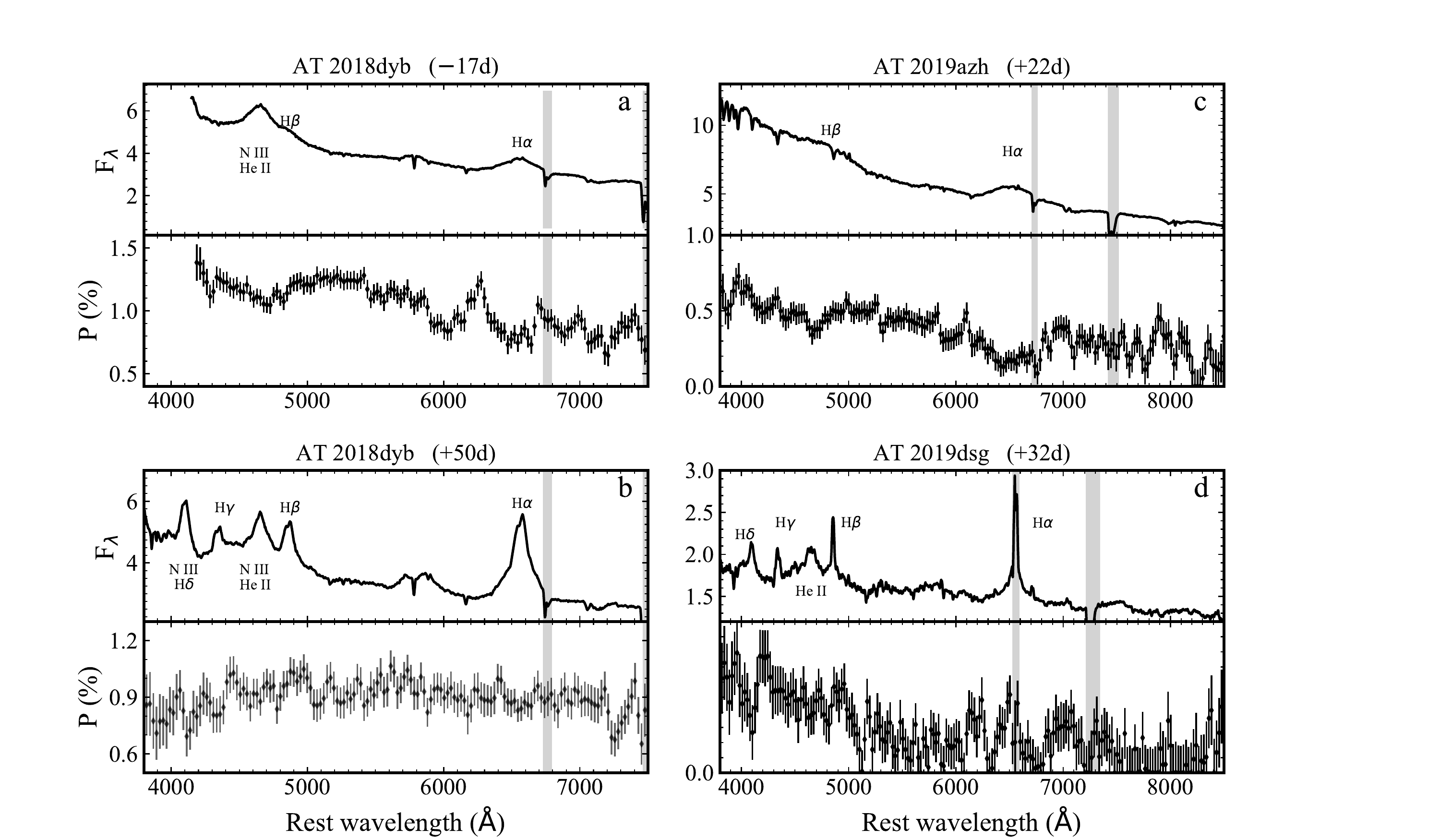}
\caption{\boldmath$\vert$\unboldmath \hspace{1em} \textbf{Spectral polarimetry of optical TDEs without the ISP and the host corrections.} 
This figure is similar to Figure~\ref{fig:flux_pola_spec} but the data is shown before  applying the ISP and the host galaxy corrections.
The polarization appears to increase towards the blue but this is primarily an effect of the host contribution being a strong function of wavelength (Extended Data Figure~\ref{fig:HostContrib}) and it is not an intrinsic property of the TDEs.  
All error bars are 1$\sigma$ uncertainties and F$_\lambda$ is in units of erg s$^{-1}$ cm$^{-2}$ $\mathrm{\AA}^{-1}$.
\label{fig:flux_pola_spec_orig}
}
\end{SIfigure}

\newpage

\begin{SIfigure}
\centering
\includegraphics[width=\textwidth]{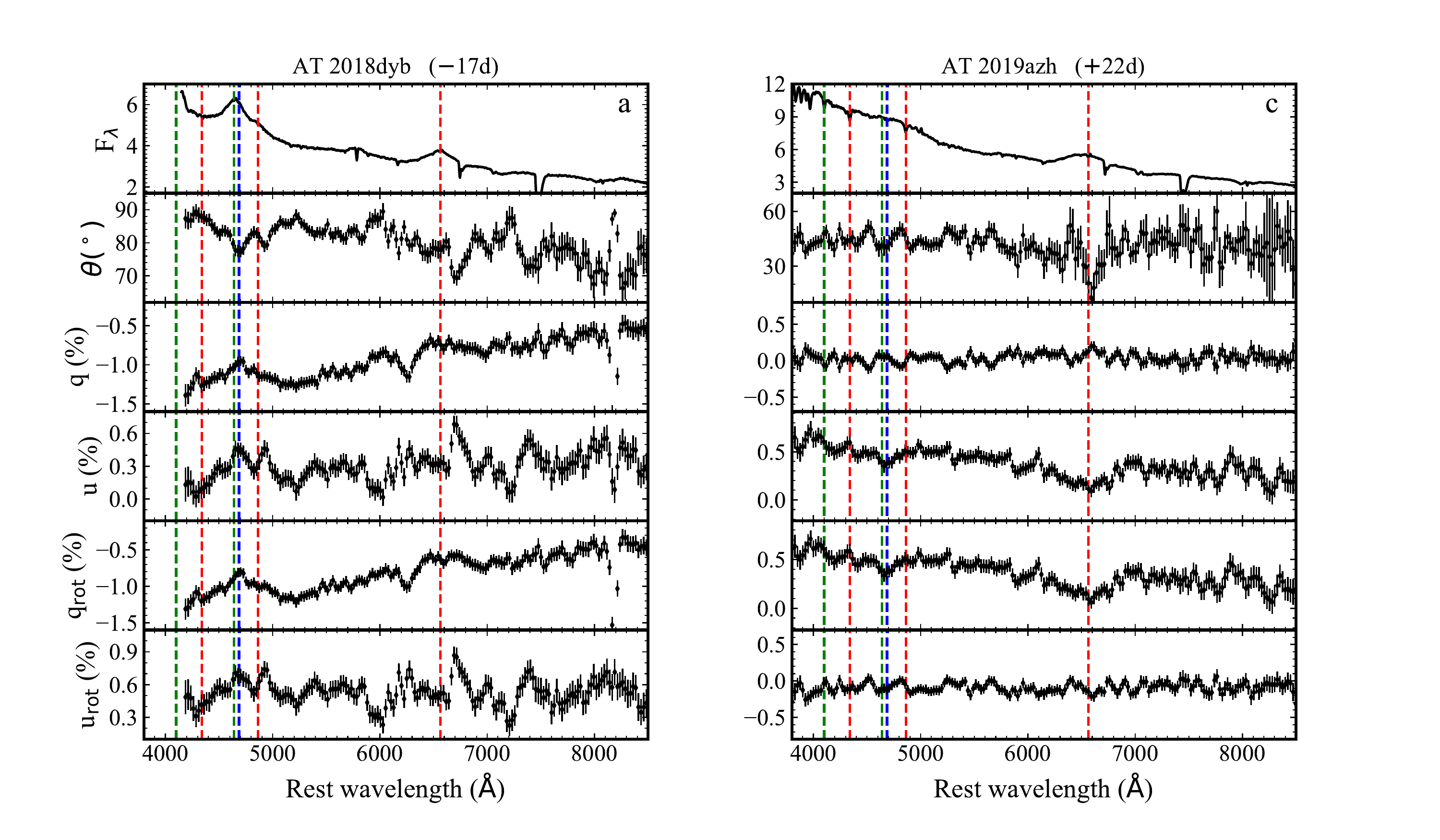}
\includegraphics[width=\textwidth]{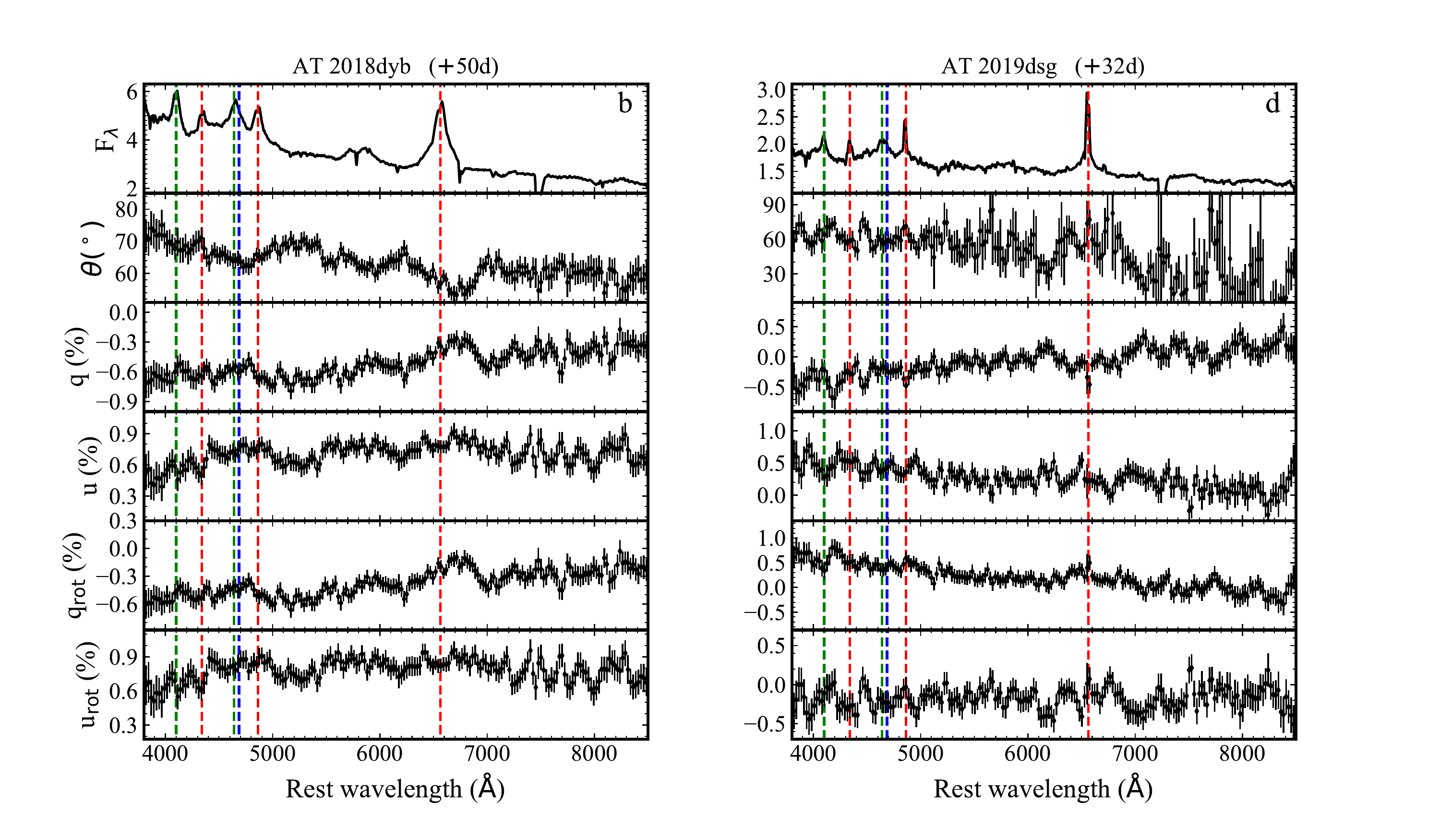}
\caption{\boldmath$\vert$\unboldmath \hspace{1em} \textbf{Polarization properties as a function of wavelength without the ISP and the host corrections.} 
This figure is similar to Figure~\ref{fig:resultsall_cor} but the data is shown before  applying the ISP and the host galaxy corrections.
All error bars are 1$\sigma$ uncertainties and F$_\lambda$ is in units of erg s$^{-1}$ cm$^{-2}$ $\mathrm{\AA}^{-1}$.
\label{fig:resultsall_orig}
}
\end{SIfigure}

\newpage

\begin{SIfigure}
\centering
\includegraphics[width=\textwidth]{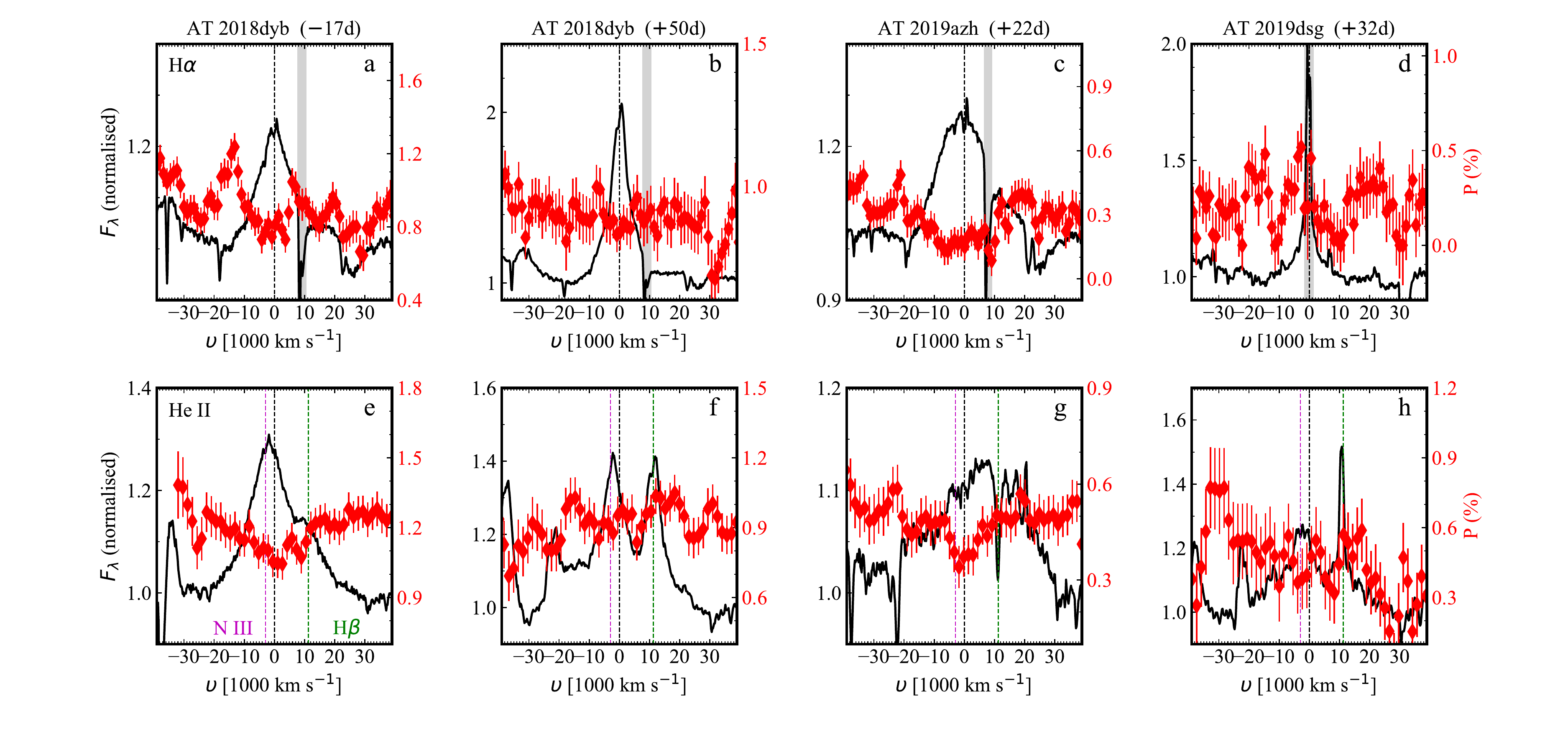}
\caption{\boldmath$\vert$\unboldmath \hspace{1em} \textbf{Emission line and polarization profiles without the ISP and the host corrections.} 
This figure is similar to Figure~\ref{fig:Ha_He_corrected} but the data is shown before  applying the ISP and the host galaxy corrections. 
All error bars are 1$\sigma$ uncertainties.
\label{fig:Ha_He_orig}
}
\end{SIfigure}

\newpage

\begin{SIfigure}
\centering
\includegraphics[width=\textwidth]{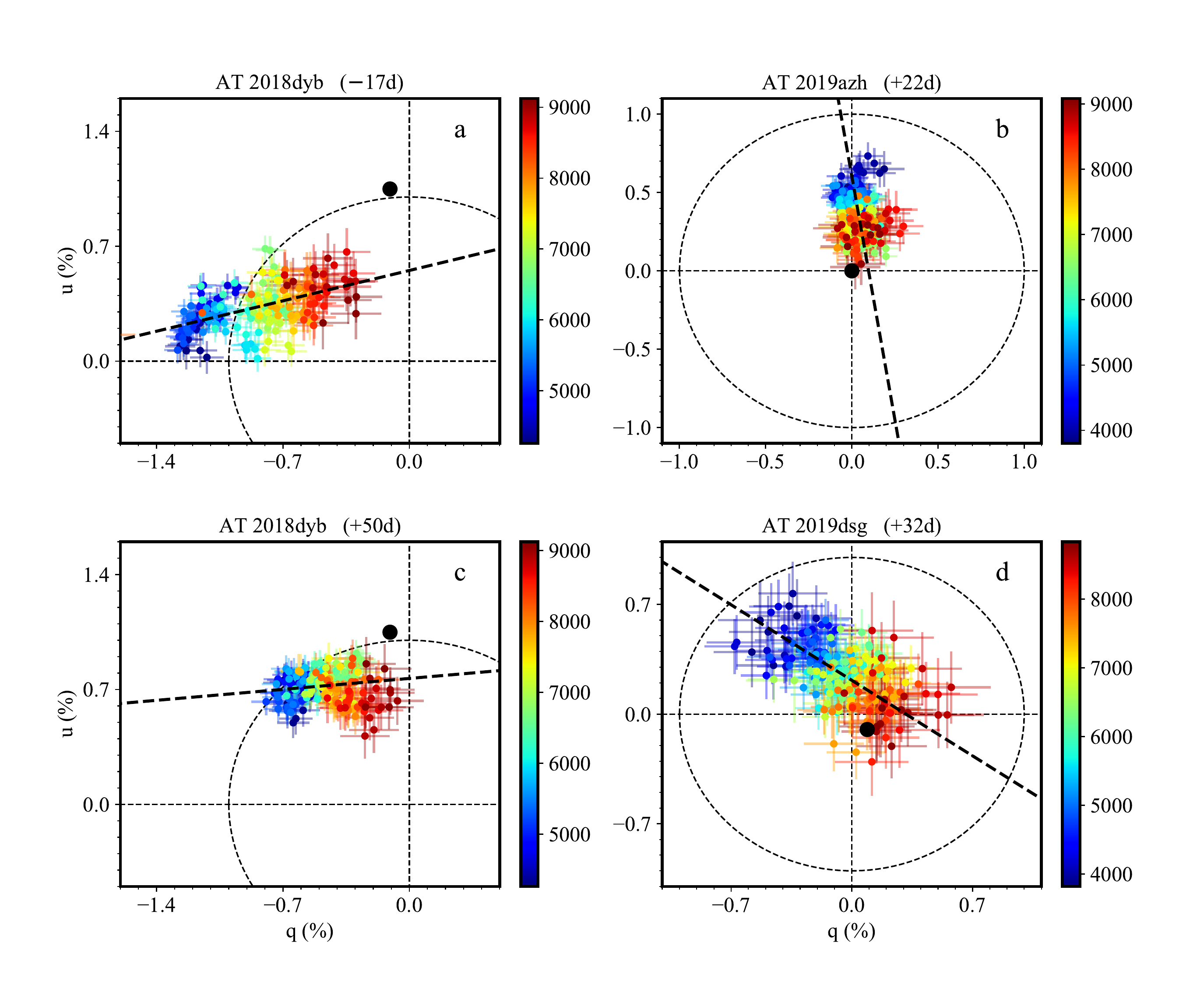}
\caption{\boldmath$\vert$\unboldmath \hspace{1em} \textbf{Stokes plane without the ISP and the host corrections.} 
This figure is similar to Figure~\ref{fig:Stokes_norot_corrected} but the data is shown before  applying the ISP and the host galaxy corrections. 
The location of the adopted ISP (see Methods) is highlighted with a black circle.
All error bars are 1$\sigma$ uncertainties.
\label{fig:Stokes_norot_original}
}
\end{SIfigure}

\newpage

\begin{SIfigure}
\centering
\includegraphics[width=16cm]{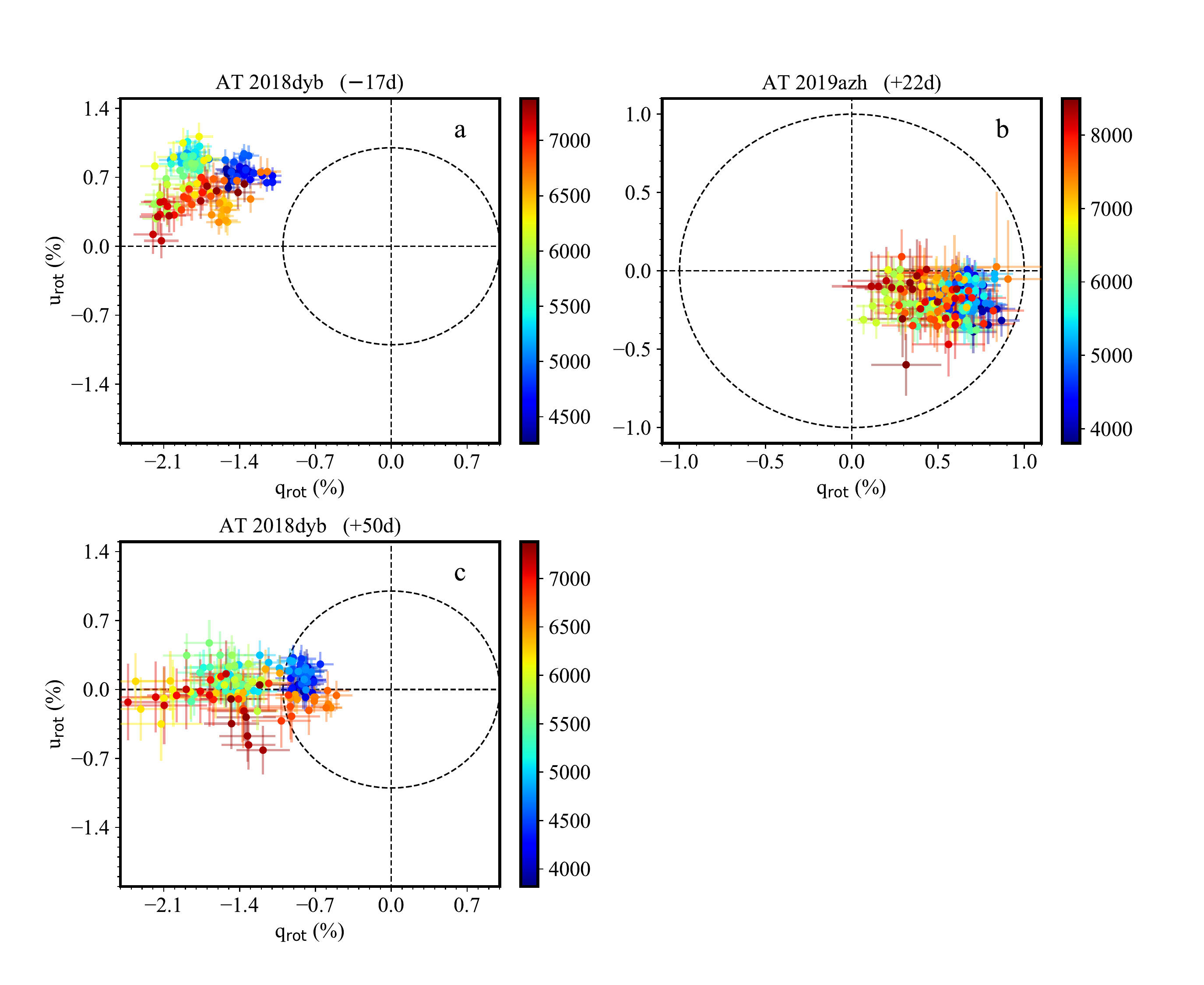}
\caption{\boldmath$\vert$\unboldmath \hspace{1em} \textbf{Rotated Stokes plane.} 
The data from Figure \ref{fig:Stokes_norot_corrected} is shown
after rotating anti-clockwise so that $q_{\mathrm{rot}}$ becomes parallel to the dominant axis. As no reliable fit was obtained for AT~2019dsg, this TDE is not shown on the rotated Stokes plane. 
We note that the pre-maximum data of AT~2018dyb shows a large systematic offset from $u_{\mathrm{rot}} = 0$, while the data at $+$50 days only shows some scatter around this axis. 
All error bars are 1$\sigma$ uncertainties.
All data is shown here after correcting for the ISP and for the host dilution.
\label{fig:Stokes_rot_corrected}
}
\end{SIfigure}

\newpage

\begin{SIfigure}
\centering
\includegraphics[width=14cm]{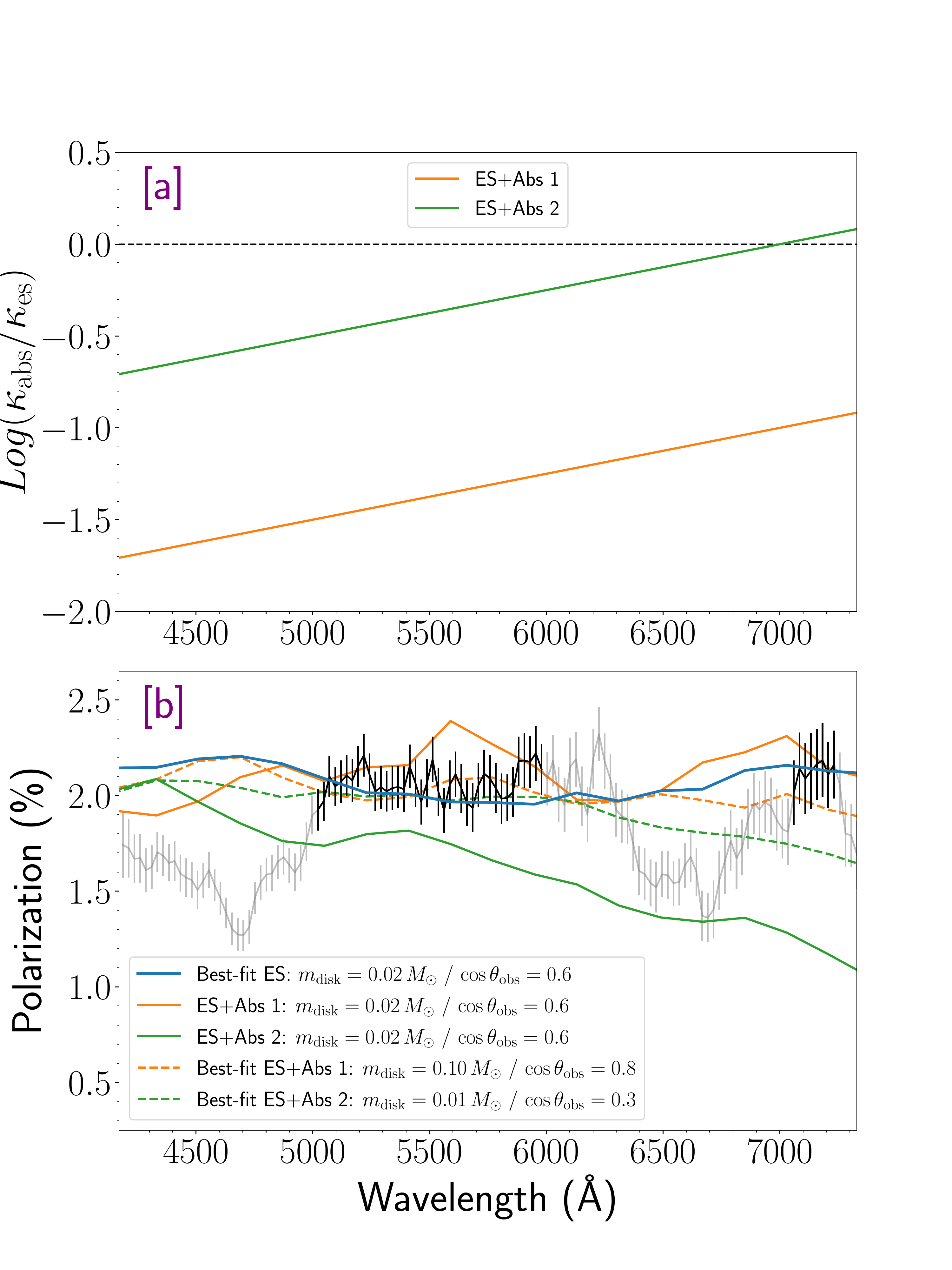}
\caption{\boldmath$\vert$\unboldmath \hspace{1em} \textbf{The impact of a depolarising absorption opacity on the wavelength dependence of polarization.} {\it Panel a}: Ratio between absorption and electron scattering opacity, $\kappa_{\rm abs}/\kappa_{\rm es}$, as a function of wavelength for Model 1 (ES+Abs 1, orange) and Model 2 (ES+Abs 2, green). {\it Panel b}: Models fit to the polarization spectrum of AT2018dyb at $-17$ days for the pure electron-scattering Model 0 (ES, cyan) and Model 1 (ES+Abs 1, orange) and Model 2 (ES+Abs 2, green) with both electron scattering and absorption opacity. See Methods for a detailed discussion. The fits are restricted to the wavelength ranges $5000-6000$ and $7050-7250$\,\AA{} (highlighted in black) that are free from strong line features in the flux spectrum. Deviations from the expect constant level in the ES model are due to Monte Carlo noise in the simulations.
All error bars are 1$\sigma$ uncertainties.
\label{fig:wavedep_a}
}
\end{SIfigure}

\newpage

\begin{SIfigure}
\centering
\includegraphics[width=\textwidth]{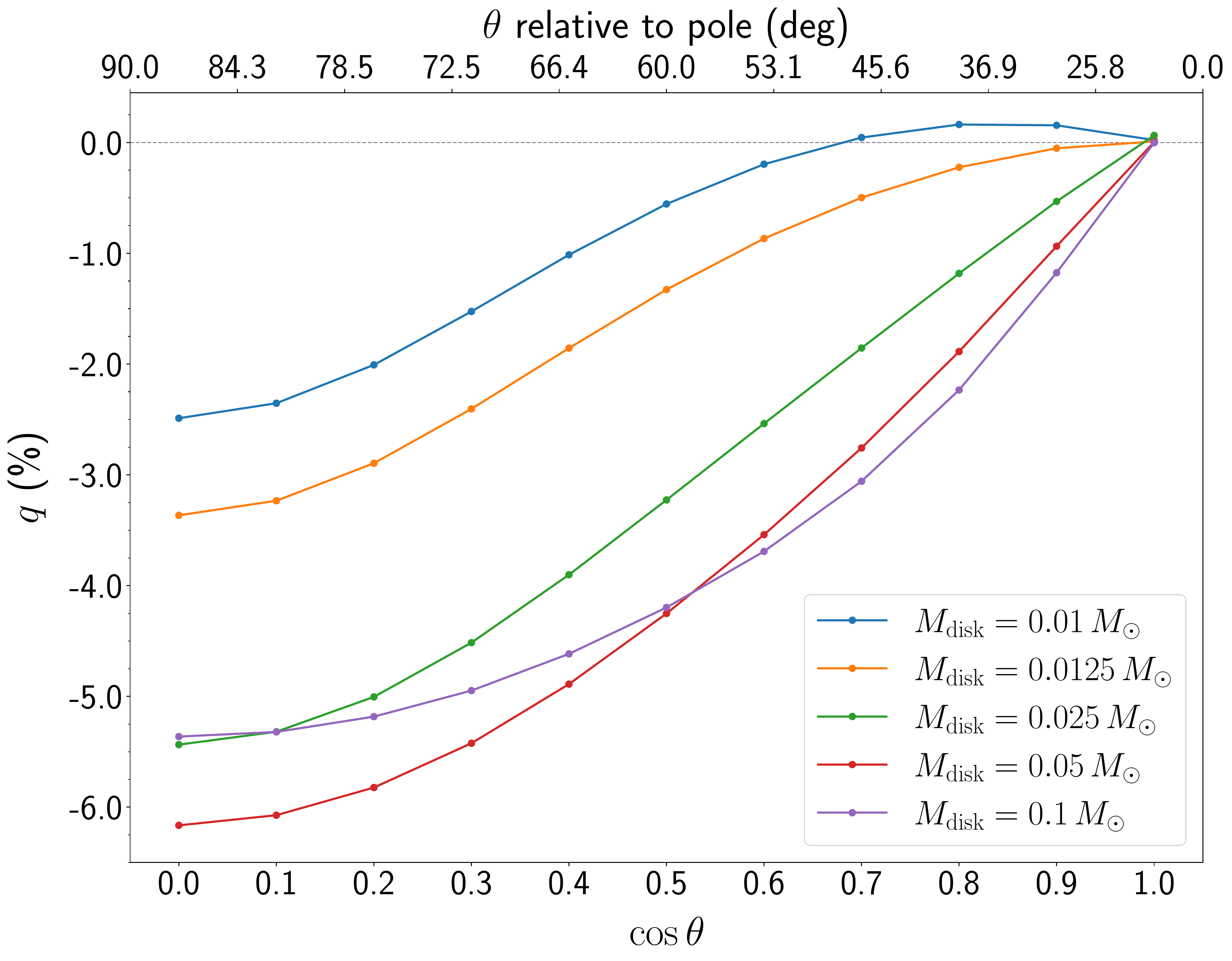}
\caption{\boldmath$\vert$\unboldmath \hspace{1em} \textbf{Polarization predictions for the TDE disk model.} The $q$ signal is shown as a function of viewing angle $\theta$. Different lines show predictions for different disk masses going from $m_\mathrm{disk}=0.01$ to $m_\mathrm{disk}=0.1\,M_\odot$. 
We observe that if $m_\mathrm{disk}$ decreases with time (as might be expected), then the polarization is also generally expected to decrease for a given viewing angle.
\label{fig:Qmodels}
}
\end{SIfigure}

\newpage

\begin{SIfigure}
\centering
\includegraphics[width=\textwidth]{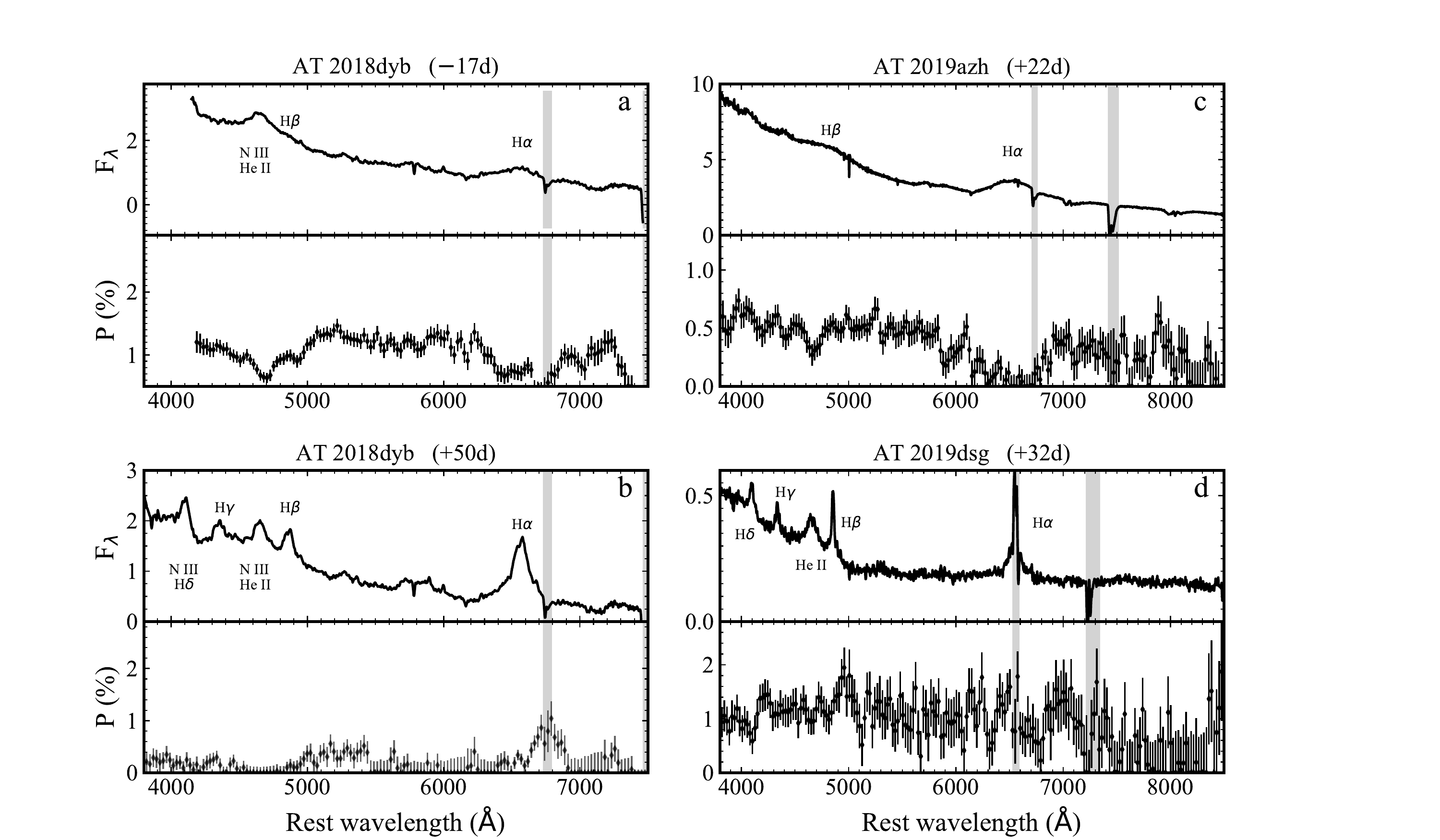}
\caption{\boldmath$\vert$\unboldmath \hspace{1em} \textbf{Spectral polarimetry of optical TDEs with alternative ISP.} 
This figure is similar to Figure~\ref{fig:flux_pola_spec} but this time we have applied an alternative solution for the ISP in AT 2018dyb and AT 2019azh (see discussion in Methods). The qualitative conclusions of the paper remain unaffected. 
All error bars are 1$\sigma$ uncertainties.
\label{fig:flux_pola_spec_altISP}
}
\end{SIfigure}


\end{addendum}

%

\clearpage

\begin{thebibliography}{10}
\expandafter\ifx\csname url\endcsname\relax
  \def\url#1{\texttt{#1}}\fi
\expandafter\ifx\csname urlprefix\endcsname\relax\def\urlprefix{URL }\fi
\providecommand{\bibinfo}[2]{#2}
\providecommand{\eprint}[2][]{\url{#2}}

\bibitem{WangWheeler}
\bibinfo{author}{{Wang}, L.} \& \bibinfo{author}{{Wheeler}, J.~C.}
\newblock \bibinfo{title}{{Spectropolarimetry of supernovae.}}
\newblock \emph{\bibinfo{journal}{\araa}} \textbf{\bibinfo{volume}{46}},
  \bibinfo{pages}{433--474} (\bibinfo{year}{2008}).

\bibitem{BullaKN}
\bibinfo{author}{{Bulla}, M.} \emph{et~al.}
\newblock \bibinfo{title}{{The origin of polarization in kilonovae and the case
  of the gravitational-wave counterpart AT 2017gfo}}.
\newblock \emph{\bibinfo{journal}{\nastro}} \textbf{\bibinfo{volume}{3}},
  \bibinfo{pages}{99--106} (\bibinfo{year}{2019}).

\bibitem{Maund2020}
\bibinfo{author}{{Maund}, J.~R.} \emph{et~al.}
\newblock \bibinfo{title}{{Polarimetry of the superluminous transient
  ASASSN-15lh}}.
\newblock \emph{\bibinfo{journal}{\mnras}} \textbf{\bibinfo{volume}{498}},
  \bibinfo{pages}{3730--3735} (\bibinfo{year}{2020}).

\bibitem{Wiersema2012}
\bibinfo{author}{{Wiersema}, K.} \emph{et~al.}
\newblock \bibinfo{title}{{Polarimetry of the transient relativistic jet of GRB
  110328/Swift J164449.3+573451}}.
\newblock \emph{\bibinfo{journal}{\mnras}} \textbf{\bibinfo{volume}{421}},
  \bibinfo{pages}{1942--1948} (\bibinfo{year}{2012}).

\bibitem{AntonucciMiller1985}
\bibinfo{author}{{Antonucci}, R.~R.~J.} \& \bibinfo{author}{{Miller}, J.~S.}
\newblock \bibinfo{title}{{Spectropolarimetry and the nature of NGC 1068.}}
\newblock \emph{\bibinfo{journal}{\apj}} \textbf{\bibinfo{volume}{297}},
  \bibinfo{pages}{621--632} (\bibinfo{year}{1985}).

\bibitem{Smith2004}
\bibinfo{author}{{Smith}, J.~E.} \emph{et~al.}
\newblock \bibinfo{title}{{Seyferts on the edge: polar scattering and
  orientation-dependent polarization in Seyfert 1 nuclei}}.
\newblock \emph{\bibinfo{journal}{\mnras}} \textbf{\bibinfo{volume}{350}},
  \bibinfo{pages}{140--160} (\bibinfo{year}{2004}).

\bibitem{Leloudas2019}
\bibinfo{author}{{Leloudas}, G.} \emph{et~al.}
\newblock \bibinfo{title}{{The Spectral Evolution of AT 2018dyb and the
  Presence of Metal Lines in Tidal Disruption Events}}.
\newblock \emph{\bibinfo{journal}{\apj}} \textbf{\bibinfo{volume}{887}},
  \bibinfo{pages}{218} (\bibinfo{year}{2019}).

\bibitem{Holoien2020}
\bibinfo{author}{{Holoien}, T. W.~S.} \emph{et~al.}
\newblock \bibinfo{title}{{The Rise and Fall of ASASSN-18pg: Following a TDE
  from Early to Late Times}}.
\newblock \emph{\bibinfo{journal}{\apj}} \textbf{\bibinfo{volume}{898}},
  \bibinfo{pages}{161} (\bibinfo{year}{2020}).

\bibitem{Hinkle2020}
\bibinfo{author}{{Hinkle}, J.~T.} \emph{et~al.}
\newblock \bibinfo{title}{{Discovery and follow-up of ASASSN-19dj: an X-ray and
  UV luminous TDE in an extreme post-starburst galaxy}}.
\newblock \emph{\bibinfo{journal}{\mnras}} \textbf{\bibinfo{volume}{500}},
  \bibinfo{pages}{1673--1696} (\bibinfo{year}{2021}).

\bibitem{Wevers2020}
\bibinfo{author}{{Wevers}, T.}
\newblock \bibinfo{title}{{Fainter harder brighter softer: a correlation
  between {\ensuremath{\alpha}}$_{ox}$, X-ray spectral state, and Eddington
  ratio in tidal disruption events}}.
\newblock \emph{\bibinfo{journal}{\mnras}} \textbf{\bibinfo{volume}{497}},
  \bibinfo{pages}{L1--L6} (\bibinfo{year}{2020}).

\bibitem{vanvelzen2021}
\bibinfo{author}{{van Velzen}, S.} \emph{et~al.}
\newblock \bibinfo{title}{{Seventeen Tidal Disruption Events from the First
  Half of ZTF Survey Observations: Entering a New Era of Population Studies}}.
\newblock \emph{\bibinfo{journal}{\apj}} \textbf{\bibinfo{volume}{908}},
  \bibinfo{pages}{4} (\bibinfo{year}{2021}).

\bibitem{sfaradi2019azh}
\bibinfo{author}{{Sfaradi}, I.} \emph{et~al.}
\newblock \bibinfo{title}{{A Late-Time Radio Flare following a Possible
  Transition in Accretion State in the Tidal Disruption Event AT 2019azh}}.
\newblock \emph{\bibinfo{journal}{arXiv:2202.00026}}  (\bibinfo{year}{2022}).

\bibitem{Stein2020}
\bibinfo{author}{{Stein}, R.} \emph{et~al.}
\newblock \bibinfo{title}{{A tidal disruption event coincident with a
  high-energy neutrino}}.
\newblock \emph{\bibinfo{journal}{\nastro}} \textbf{\bibinfo{volume}{5}},
  \bibinfo{pages}{510--518} (\bibinfo{year}{2021}).

\bibitem{Cannizzaro2021}
\bibinfo{author}{{Cannizzaro}, G.} \emph{et~al.}
\newblock \bibinfo{title}{{Accretion disc cooling and narrow absorption lines
  in the tidal disruption event AT 2019dsg}}.
\newblock \emph{\bibinfo{journal}{\mnras}} \textbf{\bibinfo{volume}{504}},
  \bibinfo{pages}{792--815} (\bibinfo{year}{2021}).

\bibitem{Lee2020}
\bibinfo{author}{{Lee}, C.-H.} \emph{et~al.}
\newblock \bibinfo{title}{{Optical Polarimetry of the Tidal Disruption Event
  AT2019DSG}}.
\newblock \emph{\bibinfo{journal}{\apjl}} \textbf{\bibinfo{volume}{892}},
  \bibinfo{pages}{L1} (\bibinfo{year}{2020}).

\bibitem{MillerAntonucci1983}
\bibinfo{author}{{Miller}, J.~S.} \& \bibinfo{author}{{Antonucci}, R.~R.~J.}
\newblock \bibinfo{title}{{Evidence for a highly polarized continuum in the
  nucleus of NGC 1068.}}
\newblock \emph{\bibinfo{journal}{\apjl}} \textbf{\bibinfo{volume}{271}},
  \bibinfo{pages}{L7--L11} (\bibinfo{year}{1983}).

\bibitem{Marin2018}
\bibinfo{author}{{Marin}, F.}
\newblock \bibinfo{title}{{Modeling optical and UV polarization of AGNs. V.
  Dilution by interstellar polarization and the host galaxy}}.
\newblock \emph{\bibinfo{journal}{\aap}} \textbf{\bibinfo{volume}{615}},
  \bibinfo{pages}{A171} (\bibinfo{year}{2018}).

\bibitem{Roth16}
\bibinfo{author}{{Roth}, N.}, \bibinfo{author}{{Kasen}, D.},
  \bibinfo{author}{{Guillochon}, J.} \& \bibinfo{author}{{Ramirez-Ruiz}, E.}
\newblock \bibinfo{title}{{The X-Ray through Optical Fluxes and Line Strengths
  of Tidal Disruption Events}}.
\newblock \emph{\bibinfo{journal}{\apj}} \textbf{\bibinfo{volume}{827}},
  \bibinfo{pages}{3} (\bibinfo{year}{2016}).

\bibitem{Bonnerot21}
\bibinfo{author}{{Bonnerot}, C.} \& \bibinfo{author}{{Stone}, N.~C.}
\newblock \bibinfo{title}{{Formation of an Accretion Flow}}.
\newblock \emph{\bibinfo{journal}{\ssr}} \textbf{\bibinfo{volume}{217}},
  \bibinfo{pages}{16} (\bibinfo{year}{2021}).

\bibitem{Wevers2019}
\bibinfo{author}{{Wevers}, T.} \emph{et~al.}
\newblock \bibinfo{title}{{Evidence for rapid disc formation and reprocessing
  in the X-ray bright tidal disruption event candidate AT 2018fyk}}.
\newblock \emph{\bibinfo{journal}{\mnras}} \textbf{\bibinfo{volume}{488}},
  \bibinfo{pages}{4816--4830} (\bibinfo{year}{2019}).

\bibitem{Inserra2016}
\bibinfo{author}{{Inserra}, C.}, \bibinfo{author}{{Bulla}, M.},
  \bibinfo{author}{{Sim}, S.~A.} \& \bibinfo{author}{{Smartt}, S.~J.}
\newblock \bibinfo{title}{{Spectropolarimetry of Superluminous Supernovae:
  Insight into Their Geometry}}.
\newblock \emph{\bibinfo{journal}{\apj}} \textbf{\bibinfo{volume}{831}},
  \bibinfo{pages}{79} (\bibinfo{year}{2016}).

\bibitem{Wiersema2020}
\bibinfo{author}{{Wiersema}, K.} \emph{et~al.}
\newblock \bibinfo{title}{{Polarimetry of relativistic tidal disruption event
  Swift J2058+0516}}.
\newblock \emph{\bibinfo{journal}{\mnras}} \textbf{\bibinfo{volume}{491}},
  \bibinfo{pages}{1771--1776} (\bibinfo{year}{2020}).

\bibitem{Roth18}
\bibinfo{author}{{Roth}, N.} \& \bibinfo{author}{{Kasen}, D.}
\newblock \bibinfo{title}{{What Sets the Line Profiles in Tidal Disruption
  Events?}}
\newblock \emph{\bibinfo{journal}{\apj}} \textbf{\bibinfo{volume}{855}},
  \bibinfo{pages}{54} (\bibinfo{year}{2018}).

\bibitem{Wang2004}
\bibinfo{author}{{Wang}, L.} \emph{et~al.}
\newblock \bibinfo{title}{{On the Hydrogen Emission from the Type Ia Supernova
  SN 2002ic}}.
\newblock \emph{\bibinfo{journal}{\apjl}} \textbf{\bibinfo{volume}{604}},
  \bibinfo{pages}{L53--L56} (\bibinfo{year}{2004}).

\bibitem{Patat2011}
\bibinfo{author}{{Patat}, F.}, \bibinfo{author}{{Taubenberger}, S.},
  \bibinfo{author}{{Benetti}, S.}, \bibinfo{author}{{Pastorello}, A.} \&
  \bibinfo{author}{{Harutyunyan}, A.}
\newblock \bibinfo{title}{{Asymmetries in the type IIn SN 2010jl}}.
\newblock \emph{\bibinfo{journal}{\aap}} \textbf{\bibinfo{volume}{527}},
  \bibinfo{pages}{L6} (\bibinfo{year}{2011}).

\bibitem{LodatoRossi}
\bibinfo{author}{{Lodato}, G.} \& \bibinfo{author}{{Rossi}, E.~M.}
\newblock \bibinfo{title}{{Multiband light curves of tidal disruption events}}.
\newblock \emph{\bibinfo{journal}{\mnras}} \textbf{\bibinfo{volume}{410}},
  \bibinfo{pages}{359--367} (\bibinfo{year}{2011}).

\bibitem{Guillochon2014}
\bibinfo{author}{{Guillochon}, J.}, \bibinfo{author}{{Manukian}, H.} \&
  \bibinfo{author}{{Ramirez-Ruiz}, E.}
\newblock \bibinfo{title}{{PS1-10jh: The Disruption of a Main-sequence Star of
  Near-solar Composition}}.
\newblock \emph{\bibinfo{journal}{\apj}} \textbf{\bibinfo{volume}{783}},
  \bibinfo{pages}{23} (\bibinfo{year}{2014}).

\bibitem{Piran2015}
\bibinfo{author}{{Piran}, T.}, \bibinfo{author}{{Svirski}, G.},
  \bibinfo{author}{{Krolik}, J.}, \bibinfo{author}{{Cheng}, R.~M.} \&
  \bibinfo{author}{{Shiokawa}, H.}
\newblock \bibinfo{title}{{Disk Formation Versus Disk Accretion -- What Powers
  Tidal Disruption Events?}}
\newblock \emph{\bibinfo{journal}{\apj}} \textbf{\bibinfo{volume}{806}},
  \bibinfo{pages}{164} (\bibinfo{year}{2015}).

\bibitem{Dai18}
\bibinfo{author}{{Dai}, L.}, \bibinfo{author}{{McKinney}, J.~C.},
  \bibinfo{author}{{Roth}, N.}, \bibinfo{author}{{Ramirez-Ruiz}, E.} \&
  \bibinfo{author}{{Miller}, M.~C.}
\newblock \bibinfo{title}{{A Unified Model for Tidal Disruption Events}}.
\newblock \emph{\bibinfo{journal}{\apj}} \textbf{\bibinfo{volume}{859}},
  \bibinfo{pages}{L20} (\bibinfo{year}{2018}).

\bibitem{thomsen2022}
\bibinfo{author}{{Thomsen}, L.~L.}, \bibinfo{author}{{Kwan}, T.},
  \bibinfo{author}{{Dai}, L.}, \bibinfo{author}{{Wu}, S.} \&
  \bibinfo{author}{{Ramirez-Ruiz}, E.}
\newblock \bibinfo{title}{{Dynamical Unification of Tidal Disruption Events}}.
\newblock \emph{\bibinfo{journal}{arXiv:2206.02804}}  (\bibinfo{year}{2022}).

\bibitem{BullaPOSSIS}
\bibinfo{author}{{Bulla}, M.}
\newblock \bibinfo{title}{{POSSIS: predicting spectra, light curves, and
  polarization for multidimensional models of supernovae and kilonovae}}.
\newblock \emph{\bibinfo{journal}{\mnras}} \textbf{\bibinfo{volume}{489}},
  \bibinfo{pages}{5037--5045} (\bibinfo{year}{2019}).

\bibitem{Gezari2021}
\bibinfo{author}{{Gezari}, S.}
\newblock \bibinfo{title}{{Tidal Disruption Events}}.
\newblock \emph{\bibinfo{journal}{\araa}} \textbf{\bibinfo{volume}{59}},
  \bibinfo{pages}{21--58} (\bibinfo{year}{2021}).

\bibitem{vanvelzen2020}
\bibinfo{author}{{van Velzen}, S.}, \bibinfo{author}{{Holoien}, T. W.~S.},
  \bibinfo{author}{{Onori}, F.}, \bibinfo{author}{{Hung}, T.} \&
  \bibinfo{author}{{Arcavi}, I.}
\newblock \bibinfo{title}{{Optical-Ultraviolet Tidal Disruption Events}}.
\newblock \emph{\bibinfo{journal}{\ssr}} \textbf{\bibinfo{volume}{216}},
  \bibinfo{pages}{124} (\bibinfo{year}{2020}).

\bibitem{charalampopoulos2022}
\bibinfo{author}{{Charalampopoulos}, P.} \emph{et~al.}
\newblock \bibinfo{title}{{A detailed spectroscopic study of tidal disruption
  events}}.
\newblock \emph{\bibinfo{journal}{\aap}} \textbf{\bibinfo{volume}{659}},
  \bibinfo{pages}{A34} (\bibinfo{year}{2022}).

\bibitem{LuBonnerot2020}
\bibinfo{author}{{Lu}, W.} \& \bibinfo{author}{{Bonnerot}, C.}
\newblock \bibinfo{title}{{Self-intersection of the fallback stream in tidal
  disruption events}}.
\newblock \emph{\bibinfo{journal}{\mnras}} \textbf{\bibinfo{volume}{492}},
  \bibinfo{pages}{686--707} (\bibinfo{year}{2020}).

\bibitem{1998Msngr..94....1A}
\bibinfo{author}{{Appenzeller}, I.} \emph{et~al.}
\newblock \bibinfo{title}{{Successful commissioning of FORS1 - the first
  optical instrument on the VLT.}}
\newblock \emph{\bibinfo{journal}{The Messenger}}
  \textbf{\bibinfo{volume}{94}}, \bibinfo{pages}{1--6} (\bibinfo{year}{1998}).

\bibitem{Patat2010}
\bibinfo{author}{{Patat}, F.} \emph{et~al.}
\newblock \bibinfo{title}{{VLT spectropolarimetry of the optical transient in
  NGC 300. Evidence of asymmetry in the circumstellar dust}}.
\newblock \emph{\bibinfo{journal}{\aap}} \textbf{\bibinfo{volume}{510}},
  \bibinfo{pages}{A108} (\bibinfo{year}{2010}).

\bibitem{Cikota2019}
\bibinfo{author}{{Cikota}, A.} \emph{et~al.}
\newblock \bibinfo{title}{{Linear spectropolarimetry of 35 Type Ia supernovae
  with VLT/FORS: an analysis of the Si II line polarization}}.
\newblock \emph{\bibinfo{journal}{\mnras}} \textbf{\bibinfo{volume}{490}},
  \bibinfo{pages}{578--599} (\bibinfo{year}{2019}).

\bibitem{Leloudas2015}
\bibinfo{author}{{Leloudas}, G.} \emph{et~al.}
\newblock \bibinfo{title}{{Polarimetry of the Superluminous Supernova LSQ14mo:
  No Evidence for Significant Deviations from Spherical Symmetry}}.
\newblock \emph{\bibinfo{journal}{\apjl}} \textbf{\bibinfo{volume}{815}},
  \bibinfo{pages}{L10} (\bibinfo{year}{2015}).

\bibitem{Leloudas2017}
\bibinfo{author}{{Leloudas}, G.} \emph{et~al.}
\newblock \bibinfo{title}{{Time-resolved Polarimetry of the Superluminous SN
  2015bn with the Nordic Optical Telescope}}.
\newblock \emph{\bibinfo{journal}{\apjl}} \textbf{\bibinfo{volume}{837}},
  \bibinfo{pages}{L14} (\bibinfo{year}{2017}).

\bibitem{Patat2006}
\bibinfo{author}{{Patat}, F.} \& \bibinfo{author}{{Romaniello}, M.}
\newblock \bibinfo{title}{{Error Analysis for Dual-Beam Optical Linear
  Polarimetry}}.
\newblock \emph{\bibinfo{journal}{\pasp}} \textbf{\bibinfo{volume}{118}},
  \bibinfo{pages}{146--161} (\bibinfo{year}{2006}).

\bibitem{FORS2manual}
\bibinfo{author}{ESO}.
\newblock \emph{\bibinfo{title}{FORS2 User Manual}}, vol.
  \bibinfo{volume}{96.0} (\bibinfo{publisher}{European Southern Observatory},
  \bibinfo{year}{2015}).

\bibitem{Wang1997}
\bibinfo{author}{{Wang}, L.}, \bibinfo{author}{{Wheeler}, J.~C.} \&
  \bibinfo{author}{{H{\"o}flich}, P.}
\newblock \bibinfo{title}{{Polarimetry of the Type IA Supernova SN 1996X}}.
\newblock \emph{\bibinfo{journal}{\apjl}} \textbf{\bibinfo{volume}{476}},
  \bibinfo{pages}{L27--L30} (\bibinfo{year}{1997}).

\bibitem{Heiles2000}
\bibinfo{author}{{Heiles}, C.}
\newblock \bibinfo{title}{{9286 Stars: An Agglomeration of Stellar Polarization
  Catalogs}}.
\newblock \emph{\bibinfo{journal}{\aj}} \textbf{\bibinfo{volume}{119}},
  \bibinfo{pages}{923--927} (\bibinfo{year}{2000}).

\bibitem{GonzalezGaitan2020}
\bibinfo{author}{{Gonz{\'a}lez-Gait{\'a}n}, S.} \emph{et~al.}
\newblock \bibinfo{title}{{Tips and tricks in linear imaging polarimetry of
  extended sources with FORS2 at the VLT}}.
\newblock \emph{\bibinfo{journal}{\aap}} \textbf{\bibinfo{volume}{634}},
  \bibinfo{pages}{A70} (\bibinfo{year}{2020}).

\bibitem{Serkowski1975}
\bibinfo{author}{{Serkowski}, K.}, \bibinfo{author}{{Mathewson}, D.~S.} \&
  \bibinfo{author}{{Ford}, V.~L.}
\newblock \bibinfo{title}{{Wavelength dependence of interstellar polarization
  and ratio of total to selective extinction.}}
\newblock \emph{\bibinfo{journal}{\apj}} \textbf{\bibinfo{volume}{196}},
  \bibinfo{pages}{261--290} (\bibinfo{year}{1975}).

\bibitem{Whittet1992}
\bibinfo{author}{{Whittet}, D.~C.~B.} \emph{et~al.}
\newblock \bibinfo{title}{{Systematic Variations in the Wavelength Dependence
  of Interstellar Linear Polarization}}.
\newblock \emph{\bibinfo{journal}{\apj}} \textbf{\bibinfo{volume}{386}},
  \bibinfo{pages}{562} (\bibinfo{year}{1992}).

\bibitem{Arcavi2014}
\bibinfo{author}{{Arcavi}, I.} \emph{et~al.}
\newblock \bibinfo{title}{{A Continuum of H- to He-rich Tidal Disruption
  Candidates With a Preference for E+A Galaxies}}.
\newblock \emph{\bibinfo{journal}{\apj}} \textbf{\bibinfo{volume}{793}},
  \bibinfo{pages}{38} (\bibinfo{year}{2014}).

\bibitem{French2016}
\bibinfo{author}{{French}, K.~D.}, \bibinfo{author}{{Arcavi}, I.} \&
  \bibinfo{author}{{Zabludoff}, A.}
\newblock \bibinfo{title}{{Tidal Disruption Events Prefer Unusual Host
  Galaxies}}.
\newblock \emph{\bibinfo{journal}{\apjl}} \textbf{\bibinfo{volume}{818}},
  \bibinfo{pages}{L21} (\bibinfo{year}{2016}).

\bibitem{Auchettl2017}
\bibinfo{author}{{Auchettl}, K.}, \bibinfo{author}{{Guillochon}, J.} \&
  \bibinfo{author}{{Ramirez-Ruiz}, E.}
\newblock \bibinfo{title}{{New Physical Insights about Tidal Disruption Events
  from a Comprehensive Observational Inventory at X-Ray Wavelengths}}.
\newblock \emph{\bibinfo{journal}{\apj}} \textbf{\bibinfo{volume}{838}},
  \bibinfo{pages}{149} (\bibinfo{year}{2017}).

\bibitem{MillerGoodrich}
\bibinfo{author}{{Miller}, J.~S.} \& \bibinfo{author}{{Goodrich}, R.~W.}
\newblock \bibinfo{title}{{Spectropolarimetry of High-Polarization Seyfert 2
  Galaxies and Unified Seyfert Theories}}.
\newblock \emph{\bibinfo{journal}{\apj}} \textbf{\bibinfo{volume}{355}},
  \bibinfo{pages}{456} (\bibinfo{year}{1990}).

\bibitem{Mckinney14}
\bibinfo{author}{{McKinney}, J.~C.}, \bibinfo{author}{{Tchekhovskoy}, A.},
  \bibinfo{author}{{Sadowski}, A.} \& \bibinfo{author}{{Narayan}, R.}
\newblock \bibinfo{title}{{Three-dimensional general relativistic radiation
  magnetohydrodynamical simulation of super-Eddington accretion, using a new
  code HARMRAD with M1 closure}}.
\newblock \emph{\bibinfo{journal}{\mnras}} \textbf{\bibinfo{volume}{441}},
  \bibinfo{pages}{3177--3208} (\bibinfo{year}{2014}).

\bibitem{Ohsuga2011}
\bibinfo{author}{{Ohsuga}, K.} \& \bibinfo{author}{{Mineshige}, S.}
\newblock \bibinfo{title}{{Global Structure of Three Distinct Accretion Flows
  and Outflows around Black Holes from Two-dimensional
  Radiation-magnetohydrodynamic Simulations}}.
\newblock \emph{\bibinfo{journal}{\apj}} \textbf{\bibinfo{volume}{736}},
  \bibinfo{pages}{2} (\bibinfo{year}{2011}).

\bibitem{Jiang2014}
\bibinfo{author}{{Jiang}, Y.-F.}, \bibinfo{author}{{Stone}, J.~M.} \&
  \bibinfo{author}{{Davis}, S.~W.}
\newblock \bibinfo{title}{{A Global Three-dimensional Radiation
  Magneto-hydrodynamic Simulation of Super-Eddington Accretion Disks}}.
\newblock \emph{\bibinfo{journal}{\apj}} \textbf{\bibinfo{volume}{796}},
  \bibinfo{pages}{106} (\bibinfo{year}{2014}).

\bibitem{Strubbe2009}
\bibinfo{author}{{Strubbe}, L.~E.} \& \bibinfo{author}{{Quataert}, E.}
\newblock \bibinfo{title}{{Optical flares from the tidal disruption of stars by
  massive black holes}}.
\newblock \emph{\bibinfo{journal}{\mnras}} \textbf{\bibinfo{volume}{400}},
  \bibinfo{pages}{2070--2084} (\bibinfo{year}{2009}).

\bibitem{Bulla2015}
\bibinfo{author}{{Bulla}, M.}, \bibinfo{author}{{Sim}, S.~A.} \&
  \bibinfo{author}{{Kromer}, M.}
\newblock \bibinfo{title}{{Polarization spectral synthesis for Type Ia
  supernova explosion models}}.
\newblock \emph{\bibinfo{journal}{\mnras}} \textbf{\bibinfo{volume}{450}},
  \bibinfo{pages}{967--981} (\bibinfo{year}{2015}).

\bibitem{Bulla2021}
\bibinfo{author}{{Bulla}, M.} \emph{et~al.}
\newblock \bibinfo{title}{{Polarized kilonovae from black hole-neutron star
  mergers}}.
\newblock \emph{\bibinfo{journal}{\mnras}} \textbf{\bibinfo{volume}{501}},
  \bibinfo{pages}{1891--1899} (\bibinfo{year}{2021}).

\bibitem{Kasen2003}
\bibinfo{author}{{Kasen}, D.} \emph{et~al.}
\newblock \bibinfo{title}{{Analysis of the Flux and Polarization Spectra of the
  Type Ia Supernova SN 2001el: Exploring the Geometry of the High-Velocity
  Ejecta}}.
\newblock \emph{\bibinfo{journal}{\apj}} \textbf{\bibinfo{volume}{593}},
  \bibinfo{pages}{788--808} (\bibinfo{year}{2003}).

\bibitem{Walker1968}
\bibinfo{author}{{Walker}, M.~F.}
\newblock \bibinfo{title}{{Studies of Extragalactic Nebulae. V. Motions in the
  Seyfert Galaxy NGC 1068}}.
\newblock \emph{\bibinfo{journal}{\apj}} \textbf{\bibinfo{volume}{151}},
  \bibinfo{pages}{71} (\bibinfo{year}{1968}).

\bibitem{Alexander2020}
\bibinfo{author}{{Alexander}, K.~D.}, \bibinfo{author}{{van Velzen}, S.},
  \bibinfo{author}{{Horesh}, A.} \& \bibinfo{author}{{Zauderer}, B.~A.}
\newblock \bibinfo{title}{{Radio Properties of Tidal Disruption Events}}.
\newblock \emph{\bibinfo{journal}{\ssr}} \textbf{\bibinfo{volume}{216}},
  \bibinfo{pages}{81} (\bibinfo{year}{2020}).

\bibitem{Zauderer2011}
\bibinfo{author}{{Zauderer}, B.~A.} \emph{et~al.}
\newblock \bibinfo{title}{{Birth of a relativistic outflow in the unusual
  {\ensuremath{\gamma}}-ray transient Swift J164449.3+573451}}.
\newblock \emph{\bibinfo{journal}{\nat}} \textbf{\bibinfo{volume}{476}},
  \bibinfo{pages}{425--428} (\bibinfo{year}{2011}).

\bibitem{Cenko2012}
\bibinfo{author}{{Cenko}, S.~B.} \emph{et~al.}
\newblock \bibinfo{title}{{Swift J2058.4+0516: Discovery of a Possible Second
  Relativistic Tidal Disruption Flare?}}
\newblock \emph{\bibinfo{journal}{\apj}} \textbf{\bibinfo{volume}{753}},
  \bibinfo{pages}{77} (\bibinfo{year}{2012}).

\bibitem{vanvelzenTDEreverb}
\bibinfo{author}{{van Velzen}, S.}, \bibinfo{author}{{Pasham}, D.~R.},
  \bibinfo{author}{{Komossa}, S.}, \bibinfo{author}{{Yan}, L.} \&
  \bibinfo{author}{{Kara}, E.~A.}
\newblock \bibinfo{title}{{Reverberation in Tidal Disruption Events: Dust
  Echoes, Coronal Emission Lines, Multi-wavelength Cross-correlations, and
  QPOs}}.
\newblock \emph{\bibinfo{journal}{\ssr}} \textbf{\bibinfo{volume}{217}},
  \bibinfo{pages}{63} (\bibinfo{year}{2021}).

\bibitem{Jiang2021}
\bibinfo{author}{{Jiang}, N.} \emph{et~al.}
\newblock \bibinfo{title}{{Infrared Echoes of Optical Tidal Disruption Events:
  {\ensuremath{\sim}}1\% Dust-covering Factor or Less at Subparsec Scale}}.
\newblock \emph{\bibinfo{journal}{\apj}} \textbf{\bibinfo{volume}{911}},
  \bibinfo{pages}{31} (\bibinfo{year}{2021}).

\bibitem{Jiang2017}
\bibinfo{author}{{Jiang}, N.} \emph{et~al.}
\newblock \bibinfo{title}{{Mid-infrared Flare of TDE Candidate PS16dtm: Dust
  Echo and Implications for the Spectral Evolution}}.
\newblock \emph{\bibinfo{journal}{\apj}} \textbf{\bibinfo{volume}{850}},
  \bibinfo{pages}{63} (\bibinfo{year}{2017}).

\bibitem{Wang2022}
\bibinfo{author}{{Wang}, Y.} \emph{et~al.}
\newblock \bibinfo{title}{{Discovery of ATLAS17jrp as an Optical-, X-Ray-, and
  Infrared-bright Tidal Disruption Event in a Star-forming Galaxy}}.
\newblock \emph{\bibinfo{journal}{\apjl}} \textbf{\bibinfo{volume}{930}},
  \bibinfo{pages}{L4} (\bibinfo{year}{2022}).
\newblock \eprint{2204.05461}.

\bibitem{GoosmannGaskell2007}
\bibinfo{author}{{Goosmann}, R.~W.} \& \bibinfo{author}{{Gaskell}, C.~M.}
\newblock \bibinfo{title}{{Modeling optical and UV polarization of AGNs. I.
  Imprints of individual scattering regions}}.
\newblock \emph{\bibinfo{journal}{\aap}} \textbf{\bibinfo{volume}{465}},
  \bibinfo{pages}{129--145} (\bibinfo{year}{2007}).

\bibitem{Zubko2000}
\bibinfo{author}{{Zubko}, V.~G.} \& \bibinfo{author}{{Laor}, A.}
\newblock \bibinfo{title}{{The Spectral Signature of Dust Scattering and
  Polarization in the Near-Infrared to Far-Ultraviolet. I. Optical Depth and
  Geometry Effects}}.
\newblock \emph{\bibinfo{journal}{\apjs}} \textbf{\bibinfo{volume}{128}},
  \bibinfo{pages}{245--269} (\bibinfo{year}{2000}).

\end{thebibliography}


\end{document}